\documentclass[%
 reprint,
bibnotes,
 amsmath,amssymb,
 aps,prb,showpacs,
]{revtex4-1}

\usepackage{graphicx}% Include figure files
\usepackage{dcolumn}% Align table columns on decimal point
\usepackage{bm}% bold math
\usepackage{color}

\begin{document}

\preprint{APS/123-QED}

\title{Raman and fluorescence contributions to resonant inelastic 
soft x-ray scattering on LaAlO$_3$/SrTiO$_3$ heterostructures}

\author{F.~Pfaff$^{1}$}
\author{H.~Fujiwara$^{2}$}
\author{G.~Berner$^{1}$}
\author{A.~Yamasaki$^{3}$}
\author{H.~Niwa$^{4}$}
\author{H.~Kiuchi$^{5}$}
\author{A.~Gloskovskii$^{6}$}
\author{W.~Drube$^{6}$}
\author{O.~Kirilmaz$^{1}$}
\author{A.~Sekiyama$^{2}$}
\author{J.~Miyawaki$^{4,7}$}
\author{Y.~Harada$^{4,7}$}
\author{S.~Suga$^{8}$}
\author{M.~Sing$^{1}$}
\author{R.~Claessen$^{1}$}

\affiliation{$^{1}$R\"ontgen Center for Complex Material
Systems (RCCM) and Physikalisches Institut, Universit\"at W\"urzburg, D-97074 W\"urzburg, Germany}
\affiliation{$^{2}$Division of Materials Physics, Graduate School of Engineering Science, Osaka University, Osaka 560-8531, Japan}
\affiliation{$^{4}$Institute for Solid State Physics, The University of Tokyo, Chiba 277-8581, Japan}
\affiliation{$^{3}$Faculty of Science and Engineering, Konan University, Kobe 658-8501, Japan}
\affiliation{$^{5}$Department of Applied Chemistry, University of Tokyo, Hongo, Bunkyo, Tokyo 113-8656, Japan}
\affiliation{$^{6}$DESY Photon Science, Deutsches Elektronen-Synchrotron, D-22603 Hamburg, Germany}
\affiliation{$^{7}$Synchrotron Radiation Research Organization, The University of Tokyo, 7-3-1 Hongo, Bunkyo-ku, Tokyo 113-8656, Japan}
\affiliation{$^{8}$Institute of Scientific $\&$ Industrial Research, Osaka University, Ibaraki, Osaka 567-0047, Japan}

\date{\today}% It is always \today, today,

\begin{abstract}
We present a detailed study of the Ti~3$d$ carriers at the interface of
LaAlO$_3$/SrTiO$_3$ heterostructures by high-resolution resonant inelastic soft
x-ray scattering (RIXS), with special focus on the roles of overlayer thickness
and oxygen vacancies. Our measurements show the existence of
interfacial Ti~3$d$ electrons already \textit{below} the
critical thickness for conductivity and an increase of the total interface
charge up to a LaAlO$_3$ overlayer thickness of 6 unit cells before it 
saturates. By
comparing stoichiometric and oxygen deficient samples we observe strong Ti~3$d$
charge carrier doping by oxygen vacancies. The RIXS data combined 
with photoelectron spectroscopy and transport measurements indicate the 
simultaneous presence of localized and itinerant charge carriers. However, it is 
demonstrated that the relative amount of 
localized and itinerant Ti~$3d$ electrons in the ground state cannot be 
deduced from the relative intensities of the Raman and fluorescence peaks in 
excitation energy dependent RIXS measurements, in contrast to previous 
interpretations. Rather, we attribute the observation of either the Raman or 
the fluorescence signal to the spatial extension of the \textit{intermediate} 
state reached in the RIXS excitation process.
\end{abstract}
\pacs{78.70.Ck, 73.20.-r, 73.40.-c}

\maketitle

%\tableofcontents

\section{\label{sec:level1}Introduction}

Complex transition metal oxides exhibit a broad spectrum of 
intrinsic functionalities such as high temperature superconductivity, colossal 
magnetoresistance, ferroelectricity etc. Artificial layered oxide structures 
made from such materials may even host novel phases not existing in the 
individual constituents. A prominent example is the
formation of a high-mobility two-dimensional electron system (2DES) at the
interface between the polar LaAlO$_3$ (LAO) and the TiO$_2$-terminated non-polar
SrTiO$_3$ (STO), if the LAO film thickness exceeds three unit cells
(uc).\cite{Ohtomo2002,Thiel2006} The 2DES is formed by Ti~$3d$ electrons on the
STO side of the interface\cite{Sing2009} and a number of fascinating
properties are reported: it can be controlled by electric field 
gating,\cite{Caviglia2008}
becomes superconducting below 300 mK,\cite{Reyren2007} and -- most strikingly --
can simultaneously display ferromagnetism\cite{Li2011, Bert2011, Kalisky2012},
most likely related to the local moments of $3d$ electrons trapped by oxygen
vacancies.\cite{Popovic2008, Chan2009, Elfimov2002, Pavlenko2012}

While the exact mechanism for 2DES formation is still under debate, there seems 
to be a wide consensus that the polar discontinuity at the LAO/STO interface 
plays a central role. A commonly discussed scenario is the electronic 
reconstruction,\cite{Nakagawa2006}
in which electronic charge is effectively transferred from the
surface to the interface at the critical LAO film thickness in order to 
compensate the electrostatic potential build-up in the LAO. Thus induced extra 
electrons are confined to the interface,
populating the otherwise empty Ti~3$d$ states.\cite{Sing2009}
However, cation defects \cite{yu2014} and oxygen 
vacancies \cite{kalabukhov2007,liu2013} have also been suggested to be the 
possible origin of the
2DES. Oxygen vacancies are known to act as electron donors in STO, and oxygen 
vacancy induced conductivity 
extending deeply into the substrate has indeed
been observed for samples grown under low oxygen
pressures.\cite{Cancellieri2010} Nonetheless, for heterostructures grown under 
higher oxygen pressures and/or post-oxidized samples the conductivity remains 
finite and confined to the interface region.\cite{Basletic2008, Sing2009} 
Despite many extensive studies several crucial
issues have remained unsolved. For instance, there are serious questions 
concerning the charge carrier dichotomy responsible for the coexistence of 
superconductivity and ferromagnetism and the electronic structure of samples 
below the critical thickness.

Such heterointerfacial materials pose an exciting technical challenge to x-ray 
and
electron spectroscopies since their use is only feasible if a 
specific interfacial contrast and a large enough probing depth can be 
accomplished at the same time. So far photoelectron spectroscopy in
the soft and hard x-ray regime has been successfully applied to various oxide
heterostructures.\cite{Sing2009,Berner2013,Schuetz22015,Cancellieri2014}
However, the need of sufficiently conductive samples to avoid 
charging and the limited probing depth preventing measurements on samples with 
overlayer thicknesses exceeding 10\,uc are major disadvantages of the 
photoemission technique. Resonant inelastic x-ray scattering (RIXS) 
is appropriate to overcome these problems. As a photon-in/photon-out method it 
is bulk-sensitive and does not require intense efforts to prepare an atomically 
clean sample surface. In addition, it allows the measurement of highly 
insulating samples. Previous studies have established that the
RIXS intensity measured at the Ti $L$ (2$p$) edge provides valuable information 
on the
total amount of interfacial Ti~3$d$ electrons in LAO/STO
heterostructures.\cite{Berner2010,Zhou2011} Interestingly, the charge carrier
concentrations determined by these early RIXS experiments -- combined with data
from hard x-ray photoelectron spectroscopy (HAXPES) -- were found to be
significantly higher than those measured by Hall effect.\cite{Berner2010} This 
observation has
been attributed to the coexistence of itinerant and localized Ti~$3d$ electrons,
which contribute both to the spectroscopic signals in RIXS and HAXPES whereas 
Hall measurements are only sensitive to the mobile 
species.\cite{Berner2010} In their study of LAO/STO
superlattices Zhou \textit{et al.} argued that the Raman and incoherent 
fluorescence signals in excitation energy dependent RIXS can be used to 
distinguish between
localized and delocalized Ti\,3$d$ carriers in the ground state.\cite{Zhou2011}

In this paper, we present the results of a detailed high resolution
RIXS study of high quality LAO/STO heterostructures, probed with 
systematically varying photon energies. In agreement with the earlier
superlattice results \cite{Zhou2011} we observe two prominent RIXS peaks
involving the interfacial Ti~3$d$ electrons, a localized
$dd$-excitation staying at a constant energy loss of around 2.3\,eV
(Raman-signal) and a fluorescence signal being almost independent of
excitation energy. To further elucidate the origin of these two RIXS features we
analyze their behavior as function of film thickness and oxygen stoichiometry 
together with complementary HAXPES and Hall-effect data. Importantly, we find
that the Raman {\it vs.} fluorescence character of the RIXS peaks
reflects the different degree of localization of the \textit{intermediate}
state in the RIXS process but not that of the initial state. We are thus led to
conclude that RIXS cannot directly distinguish localized
from itinerant Ti~3$d$ carriers in STO-based heterostructures, contrary to 
earlier suggestions. \cite{Zhou2011}

 \begin{figure}
	\includegraphics[width=0.49\textwidth]{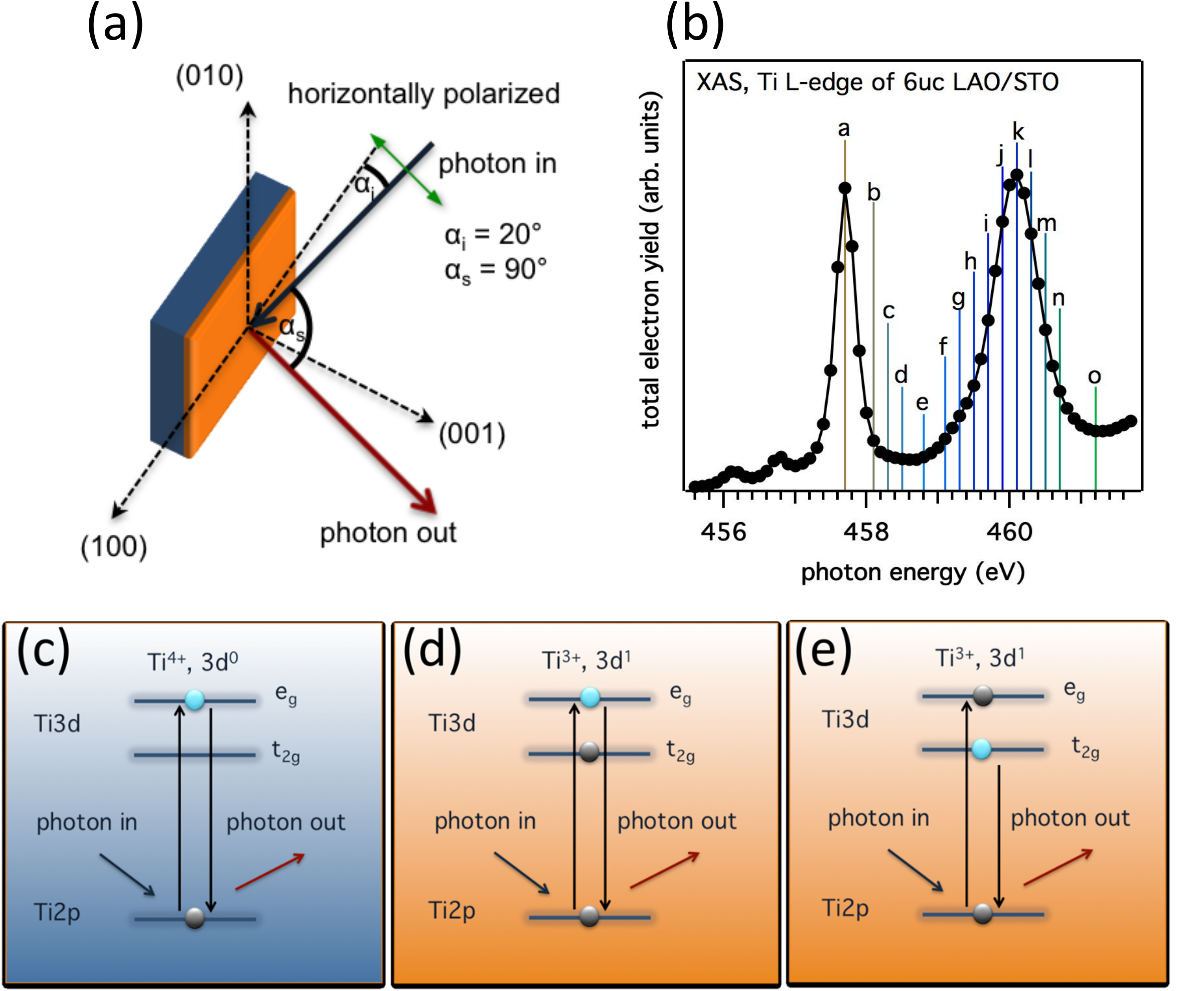}
	\caption{(a) Sketch of the experimental geometry for XAS and RIXS 
measurements. Samples were aligned with the sample normal (001) in the 
scattering plane. The incident angle of the photons was set to 20$^\circ$ from 
the sample surface while the scattering angle was always kept at 90$^\circ$. 
The polarization vector of the incoming light was set to be within the 
scattering plane (p-polarization). (b) X-ray absorption spectrum 
at the Ti $L_3$ (Ti~2$p_{3/2}$) edge of a 6\,uc LAO/STO heterostructure 
measured in total electron yield mode. The used photon energies for RIXS 
measurements are marked by labels $a$-$o$. (c-e) Basic principle of the RIXS 
process using photons tuned to, e.g., the $e$$_g$-resonance. Blue electron is 
assigned to the electron taking part in the deexcitation process (c) No 
inelastic signal 
is observable for Ti ions in a Ti~3$d$$^0$ configuration (Ti$^{4+}$). (d-e) An 
additional decay channel is possible for Ti ions in a Ti~3$d$$^1$ configuration 
(Ti~$^{3+}$), ending in an excited state with an electron in the 
$e_g$-manifold, and observable as an energy loss peak in the RIXS spectra.}
	\label{fig1}
\end{figure}

\section{\label{sec:level2}Experiment}
A series of LAO/STO heterostructures with LAO overlayer thicknesses from 2 to 
20\,uc were grown by pulsed laser deposition (PLD) on TiO$_2$-terminated 
STO(001) substrates. Laser ablation of the single crystalline LAO target was 
achieved with a KrF excimer laser at a fluency of 1.1\,J/cm$^2$ and a 
target-substrate distance of 56 mm. During the film growth the STO substrate 
was 
heated up to 780$^\circ$C and the oxygen partial pressure was set to 
1\,x\,10$^{-3}$\,mbar.
In order to suppress unwanted oxygen vacancies some samples were treated by a 
post-oxidation (PO) procedure immediately after the growth (stepwise cooling 
down from growth temperature to 360$^\circ$C in a high oxygen pressure of 
500\,mbar over a period of 2 hours). Additional samples with a film thickness of 
6\,uc and intentionally reduced oxygen stoichiometry were grown by skipping 
the post-growth oxidation step (medium pressure growth\,=\,MP). To achieve even 
higher oxygen vacancy concentrations the oxygen partial pressure during 
the growth was reduced down to 5\,x\,10$^{-7}$\,mbar for other samples (low 
pressure growth\,=\,LP). Reference measurements were performed on 
commercially available bare STO crystals.

The RIXS and x-ray absorption (XAS) measurements were performed at BL07LSU of 
SPring-8. A detailed description of the experimental setup can be found in Ref. 
\onlinecite{Harada2012}. All measurements were carried out at room temperature. 
The experimental geometry for both, RIXS and XAS, is depicted in 
Fig.~\ref{fig1}~(a). The angle of x-ray incidence was kept at 20$^\circ$ with 
respect to the sample surface while the scattering angle was always set to 
90$^\circ$ with respect to the incoming beam. The x-ray polarization was chosen 
to lie within the scattering plane ($p$-polarization). By fitting the elastic 
line of bare STO with a Gaussian the energy resolution was determined to be 
$\approx$\,90\,meV.
In order to reduce photon-induced oxygen depletion \cite{Walker2014} the photon 
spot ($\sim 2\,{\rm \mu m} \times 90\,{\rm \mu m}$) was moved across 
the sample surface by 
10\,$\mu$m every 30 minutes. No significant changes were observed in the 
spectra within this timescale.

A Ti $L_3$ edge XAS spectrum of a PO sample with an overlayer thickness of 
6\,uc measured in total electron yield mode is shown in Fig.~\ref{fig1}~(b). The 
photon energies selected for the RIXS measurements are marked by labels $a$ to 
$o$. The two peaks centered at energies $a$ and $k$ reflect the cubic 
crystal-field splitting of the Ti~3$d$ states in the STO substrate and are 
assigned to transitions from the Ti~2$p$ core level into the 3$d_{t_{2g}}$ and 
3$d_{e_g}$ states, respectively.

RIXS is a two-step process involving a photoexcitation of a core electron into 
unoccupied conduction band states and a subsequent decay of the excited state by 
emitting an x-ray photon. At the Ti $L$ edge, after the excitation 
of an electron from the Ti~2$p$ to the Ti~3$d$ shell, different radiative decay 
channels become possible resulting in final states with different electronic 
configurations. The emitted photon can be elastically scattered where the final 
state is the same as the original ground state [Figs.~\ref{fig1}(c) and 
(d)]. Alternatively, an inelastic signal is observed when the system remains in 
an excited final state with, e.g., one electron in an $e_g$-state and no 
electron in the $t_{2g}$ manifold as depicted in Fig.~\ref{fig1}(e). However, 
the latter process can only occur, if there is already one $d$-electron 
in the initial state. Consequently, such a Raman signal is a measure of the 
finite 3$d$ electron occupation as has already been shown for other titanates 
\cite{Ulrich2008,Higuchi1999} and also for LAO/STO heterostructures 
\cite{Berner2010} and superlattices.\cite{Zhou2011}

Complementary photoemission experiments have been performed at the HAXPES 
endstation of beamline P09 at PETRA III (DESY, Hamburg), equipped with a SPECS 
Phoibos 220 spectrometer. At the used photon energy of 3.5\,keV an overall 
energy resolution of 450\,meV was realized. A detailed description of the 
beamline can be found in Ref.~\onlinecite{Strempfer2013}.

\section{\label{sec:level3}Results}
\subsection{\label{sec:level31}Excitation energy dependence}
\begin{figure}[h]
	\includegraphics[width=0.49\textwidth]{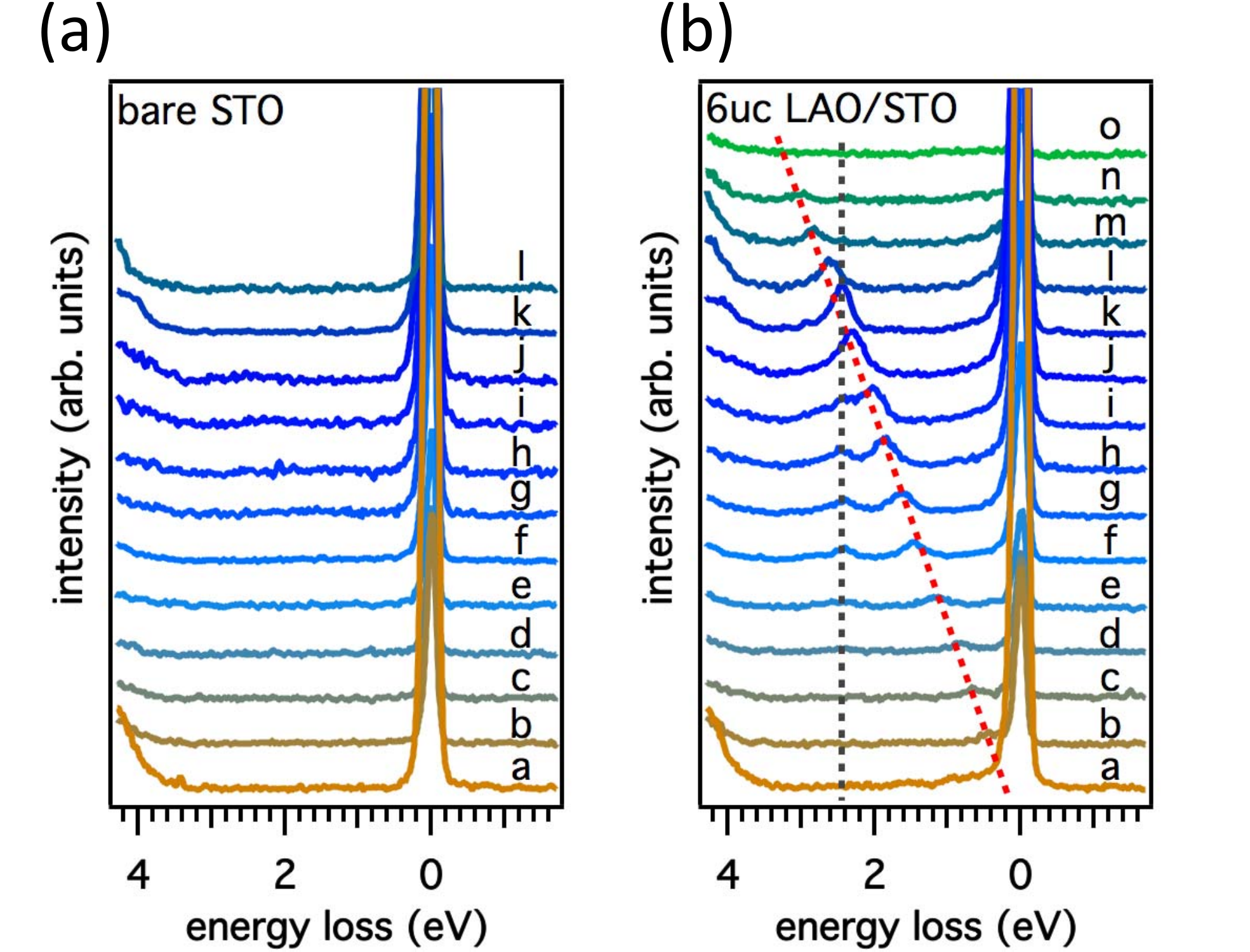}
	\caption{Series of RIXS spectra. The labels $a$-$l$ and $a$-$o$ in (a) 
and (b) correspond to the excitation energies marked in the XAS spectrum in 
Fig.~\ref{fig1}~(b). (a) Series of RIXS spectra of a bare STO sample excited 
across the Ti $L$ edge. The spectra are normalized by acquisition time. (b) 
Series of RIXS spectra of a 6\,uc LAO/STO heterostructure excited across the Ti 
$L$ edge. The spectra are normalized by acquisition time. The black and orange 
dashed line are guides to the eye showing the Raman- and the fluorescence-like 
spectral features, respectively.}
	\label{fig2}
\end{figure}

A series of RIXS spectra of a bare STO substrate as well as a post-oxidized 
LAO/STO heterostructure with a 6\,uc thick film, recorded at photon 
energies across the Ti $L$ edge and normalized to acquisition time, is 
displayed in Figs.~\ref{fig2}~(a) and (b), respectively. The labels $a$-$o$ 
correspond to the excitation photon energies marked in the absorption spectrum 
of Fig.~\ref{fig1}~(b). For the bare STO sample the final state can be either 
$2p^63d^0$ or $2p^63d^1$\underline{$L$} with 
\underline{$L$} denoting a hole in the ligand O~2$p$ shell. Beside the elastic 
line and the onset of the 3$d^1$\underline{$L$} charge transfer excitations at 
an energy loss of $\approx$\,3.5\,eV [the full energy range from 3.5 to 18.0\,eV 
energy loss is depicted in Figs.~\ref{fig2_2}~(a) and \ref{fig3}~(a)], no clear 
additional signal can be discerned throughout the photon energy series as shown 
in Fig.~\ref{fig2}~(a). In LAO/STO samples above the critical thickness the 
ground state is partially of type $2p^63d^1_{t_{2g}}$ due to the 
extra electrons in the 3$d$ shell at the conducting interface.
This results in an additional contribution to the RIXS spectra depending on the 
used excitation energy. In the case of the t$_{2g}$ resonance 
(label $a$) the excited intermediate state will have the 
configuration $2p^53d^2_{t_{2g}}$. Two different final states are possible 
either contributing to the elastic line ($2p^63d^1_{t_{2g}}$) or to the charge 
transfer excitations with an energy loss above 3.5\,eV as seen in 
Figs.~\ref{fig2_2}~(a) and \ref{fig3}~(a) and represented by 
$2p^63d^2_{t_{2g}}\underline{L}$).

Increasing the incoming photon energy to the $e$$_g$-resonance (label $k$),  
$2p^5t^1_{2g}e^1_g$ intermediate states can result via an additional decay 
channel in excited final states of the type $2p^63d^1_{e_g}$ [see 
Fig.~\ref{fig1}(e)]. Thus, an inelastic signal due to intra-atomic 
$dd$-excitations from $t_{2g}$ to $e_g$ states becomes visible at an energy 
loss of around 2.3\,eV, providing spectroscopic sensitivity to the interfacial 
Ti~3$d$ carriers [Fig.~\ref{fig2}~(b)]. When changing the excitation energy from 
the $t$$_{2g}$- towards the $e$$_g$-resonance we can identify two different 
types of signals overlapping near the $e$$_g$-resonance. While one signal shows 
a constant Raman-shift of approximately 2.3\,eV, the second signal shows 
fluorescence-like behavior with its energy position being almost independent of 
the incident photon energy. Similar results have already been obtained 
by performing excitation energy dependent RIXS on LAO/STO superlattices 
\cite{Zhou2011} and other oxides like BaTiO$_3$ and BaSO$_4$.\cite{Yoshii2012} 
In the case of the superlattices, the authors attributed the observation of a 
Raman-signal to localized charge carriers in the initial state whereas the 
fluorescence-like signal was assigned to delocalized electrons in the 
system.\cite{Zhou2011} Contrary to this interpretation, the authors of 
Ref.~\onlinecite{Yoshii2012} assign the observation of Raman-scattering in 
BaTiO$_3$ and BaSO$_4$ to a process where the excited electron is promoted to 
an unoccupied state and remains localized at the same atomic site on the 
timescale of the deexcitation, while the appearance of a fluorescence-like 
signal is ascribed to the more delocalized character of the state reached by the 
excited electron in the intermediate state. Thus further detailed experiments 
are required to clarify the origin of the features seen in the RIXS spetcra of 
LAO/STO. 

\subsection{\label{sec:level32}Thickness dependence}
\begin{figure}[t]
	\includegraphics[width=0.49\textwidth]{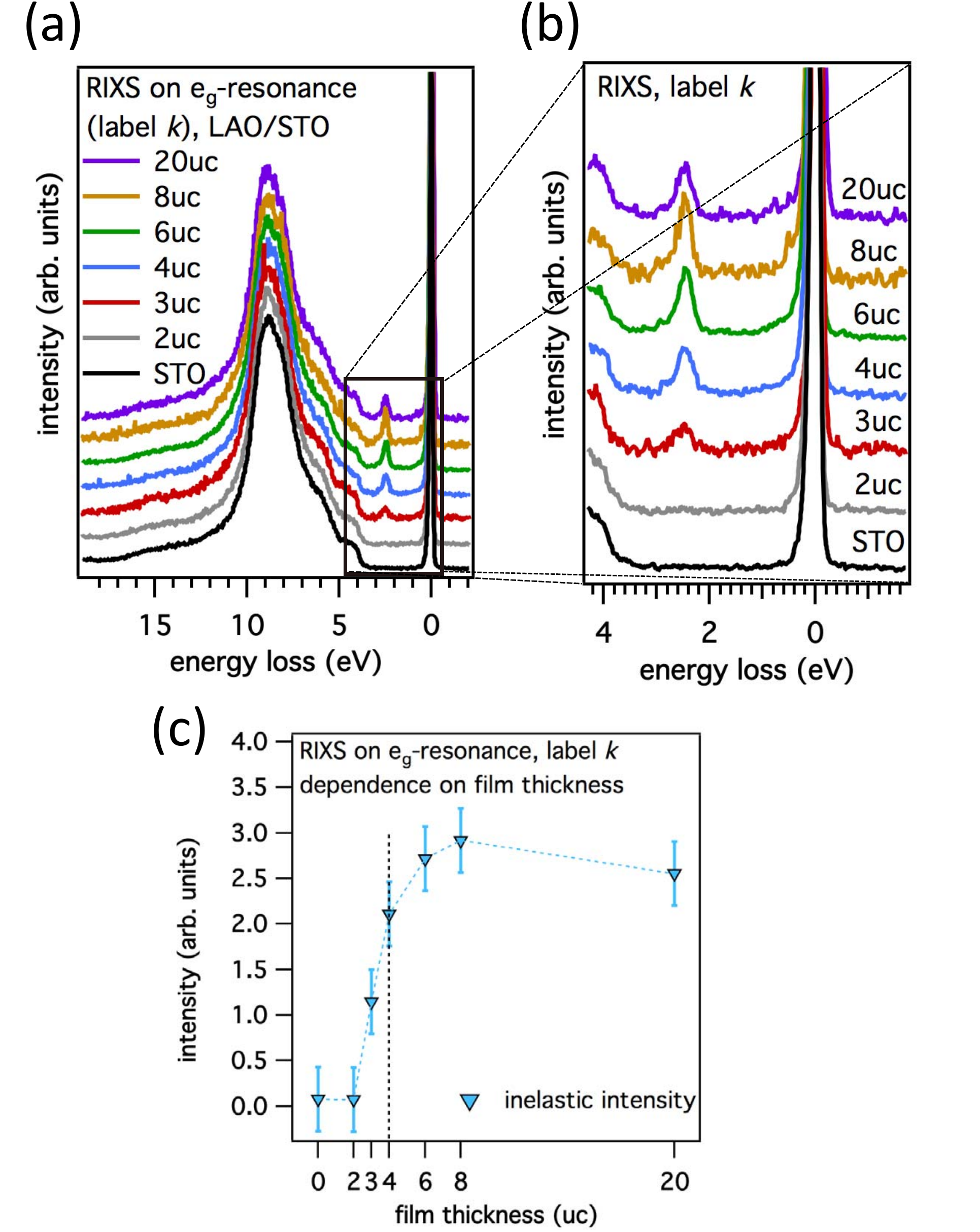}
	\caption{(a) RIXS 
spectra measured at the $e_g$-resonance of a bare STO sample and LAO/STO 
heterostructures with various overlayer thickness. The spectra have been 
normalized to the $3d^1\underline{L}$ charge transfer excitations. (b) 
Low energy loss region of the RIXS spectra in Fig.~\ref{fig2_2}~(a). (c) 
Spectral weight of the loss structure around 2.3\,eV in Figs.~\ref{fig2_2}~(a) 
and (b) from a fitting analysis plotted against the film thickness.}
	\label{fig2_2}
\end{figure}

To further elucidate the nature of the two RIXS features, we performed 
detailed RIXS measurements at the $e$$_g$-resonance on post-oxidized (PO) 
samples with overlayer thicknesses between 2 and 20\,uc as well as on bare STO 
as reference. The results obtained for this series are displayed in 
Figs.~\ref{fig2_2}~(a) and (b) showing a strong dependence on the LAO thickness. 
The data are normalized to the intensity of the 3d$^1$$\underline{L}$ charge 
transfer excitations with energy losses above 3.5\,eV which originate mainly 
from the STO substrate and therefore should be essentially constant for the 
whole set of samples. To further analyze the spectra, the RIXS signals at 
a loss energy of about 2.3 eV (see Fig.~\ref{fig2_2}~(b)) have been 
fitted by Gaussians and the obtained intensities plotted as a function of 
overlayer thickness in Fig.~\ref{fig2_2}~(c). The absence of any 
inelastic intensity for bare STO and the subcritical 2\,uc sample shows the lack 
of Ti sites with 3$d^1$-configuration (i.e., with extra electrons in the 
$d$-shell in the initial state).
Although the photon-induced generation of oxygen vacancies, i.e., extrinsic 
electron doping, is minimized by moving the sample continuously during the 
measurements, spectral weight is clearly visible for the insulating 3\,uc 
sample. Since we do not observe any RIXS signal for the insulating 2\,uc 
sample we consider the observation of an inelastic signal for 3\,-\,20\,uc 
samples intrinsic to the measured heterostructure. Discrepancies between the 
present and previous RIXS measurements, which do show finite inelastic spectral 
weight for the insulating 2\,uc heterostructure, likely can be attributed to a 
small amount of oxygen vacancies. These are caused by the intense synchrotron 
radiation and become observable if the same sample spot is constantly exposed 
to the beam as in the previous study\cite{Berner2010}.

The RIXS intensity for the insulating 3\,uc sample in the present experiment 
(see Fig.~\ref{fig2_2}~(b)) can be explained by the presence of $localized$ 
electrons that are not due to oxygen vacancy doping. This is in line with 
photoemission spectroscopy measurements showing indications for an incipient 
insulator-metal transition, as signalled by the appearance of Ti~3$d$ carriers 
before interface conductivity is observed.\cite{Takizawa2011} Results of x-ray 
absorption for an insulating 2\,uc sample that reveal an orbital 
reconstruction attributable to interface symmetry breaking and the transfer of 
localized electrons to interface states, support this 
scenario.\cite{Salluzzo2013} Our findings are also in line with the critical 
thickness of 3\,uc for ferromagnetism --- probably due to the existence of local 
Ti~3$d$ moments --- observed by Kalisky $et$ $al.$ in overlayer thickness 
dependent scanning SQUID measurements.\cite{Kalisky2012}

The observed RIXS intensity increases continuously with LAO film thickness and 
saturates at a thickness of 6\,uc. In contrast, the mobile 
charge carrier concentration obtained from Hall measurements shows a sharp 
step-like behavior at an overlayer thickness of 4\,uc\cite{Thiel2006}. This 
discrepancy has been attributed in previous experimental and theoretical studies 
to the presence of delocalized and localized charge carriers where the latter 
are not observed in transport.\cite{Thiel2006,Popovic2008,Son2009,Berner2010}

\begin{figure}
	\includegraphics[width=0.49\textwidth]{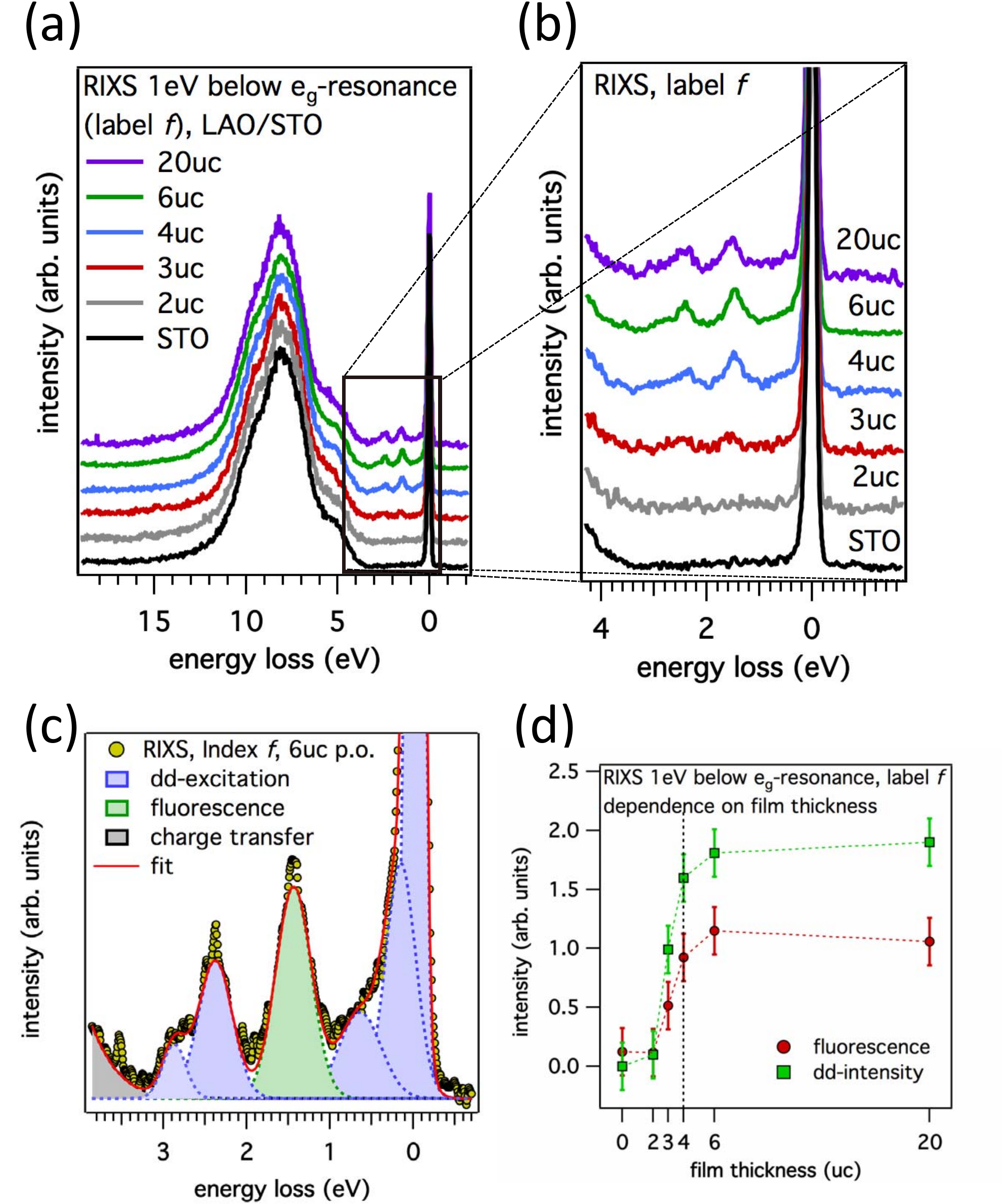}
	\caption{(a) RIXS spectra of bare STO and LAO/STO heterostructures with 
varying LAO overlayer thickness measured 1\,eV below the $e$$_g$-resonance. The 
spectra have been normalized to the 3$d$$^1$$\underline{L}$ charge transfer 
excitations. (b) Close-up of the region marked by the rectangle 
in Fig.~\ref{fig3}~(a). (c) Fit analysis of the off-resonance RIXS spectra of 
the 6\,uc LAO/STO sample using 7 Gaussian peaks. (d) Inelastic spectral weight 
of the $dd$-excitations and the fluorescence-like feature obtained from 
the fitting of the RIXS spectra in Fig.~\ref{fig3}~(a) and (b) plotted as 
function of LAO overlayer thickness.}
	\label{fig3}
\end{figure}

Since the inelastic signal at the $e$$_g$-resonance is a superposition of two 
different features as can be recognized from Fig.~\ref{fig2}~(b), we focus in 
the following on measurements using an excitation energy 1\,eV below the 
$e$$_g$-resonance (label $f$), where the Raman- and fluorescence-like components 
are well separated. The spectra recorded under this condition are depicted in 
Fig.~\ref{fig3}~(a) and (b) for the full set 
of samples. As selectively shown for the 6\,uc sample in 
Fig.~\ref{fig3}~(c), the spectrum in this region contains seven components 
including the contribution of the charge transfer excitations at energies 
higher than 3.5 eV. In addition to the elastic line at zero loss energy the 
fluorescence-like component (green shaded) at 1.45\,eV and four 
distinct $dd$-excitations (purple shaded) can be distinguished 
owing to the high energy resolution. The latter observation indicates the 
complete lifting of the Ti~3$d$ degeneracy. By careful fitting of the data 
with Gaussian line shapes, the energies of the $dd$-transitions on the 
Ti$^{3+}$ 
ions are determined to be 0.17, 0.61, 2.38, and 2.88\,eV with an accuracy of 
$\pm$\,0.10\,eV. In a one-particle picture, from these excitation energies 
the energy separation of the lowest and the highest $t_{2g}$-orbital can be 
estimated as (0.60\,$\pm$\,0.10)\,eV and the $e$$_g$-splitting as 
(0.50\,$\pm$\,0.10)\,eV. These values significantly exceed those determined 
from x-ray absorption for the interface Ti$^{4+}$-ions.\cite{Salluzzo2009} 
RIXS measurements of $dd$ excitation energies on LAO/STO superlattices, again 
representative for the (distorted) Ti$^{3+}$O$_6$ octahedra at the interface,  
yield values for these energy splittings comparable to ours \cite{Zhou2011}. 
The larger values, i.e., the larger crystal-field strength as compared with 
Ti$^{4+}$O$_6$ octahedra, were attributed in this study to an enhanced Coulomb 
repulsion and covalency between Ti~3$d$ and O~2$p$ states due 
to the extra occupation with Ti~3$d$ electrons.\cite{Zhou2011}

For a quantitative analysis of the intensities of the inelastic signals and 
their ratios, the spectra of the whole series of samples have been fitted 
taking seven Gaussian profiles into account as already explained in 
Fig.~\ref{fig3}~(c). 
Four represent the crystal-field split $dd$-excitations, one the 
fluorescence-like feature, and another one the elastic line. The last one
models the increasing slope of the 3$d$$^1$\underline{$L$} charge transfer 
excitations setting in at an energy loss of 3.5\,eV. The total Raman 
intensity and the intensity of the fluorescence peak are plotted as a 
function of film thickness in Fig.~\ref{fig3}~(d). As all four $dd$-excitations 
display the same dependence on LAO thickness within experimental accuracy their 
intensities have been summed up for clarity and better statistics.

Except for little spectral weight which we ascribe to 
radiation induced oxygen vacancies, no RIXS intensity is visible for bare STO 
and the 2\,uc LAO/STO sample within the accuracy of our experiment. The total 
RIXS spectral weight increases for larger overlayer thicknesses and saturates 
for a film thickness larger than 6\,uc. Thereby, the intensity ratio of the 
fluorescence-like signal and the $dd$-excitations stays essentially constant. 
The parallel rise in intensity of both signals up to 6\,uc in contrast to the 
step-like increase of the mobile charge carrier concentration at 4\,uc as 
observed in Hall measurements invalidates the attribution of either of the two 
observed features to $delocalized$ ground state electrons in LAO/STO 
heterostructures. Also the finite Raman- $and$ fluorescence-like spectral weight 
for the non-conducting 3\,uc sample contradicts such an interpretation. 
Furthermore, since both structures show essentially the same dependence on film 
thickness they rather seem to reflect the total charge carrier concentration, 
including contributions by photogenerated charge carriers, as has been pointed 
out before for RIXS measurements on the $e_g$-resonance where Raman 
and fluorescence peak overlap.\cite{Berner2010}

\subsection{\label{sec:level33}Influence of oxygen vacancies: electron doping}
\begin{figure}[h]
	\includegraphics[width=0.49\textwidth]{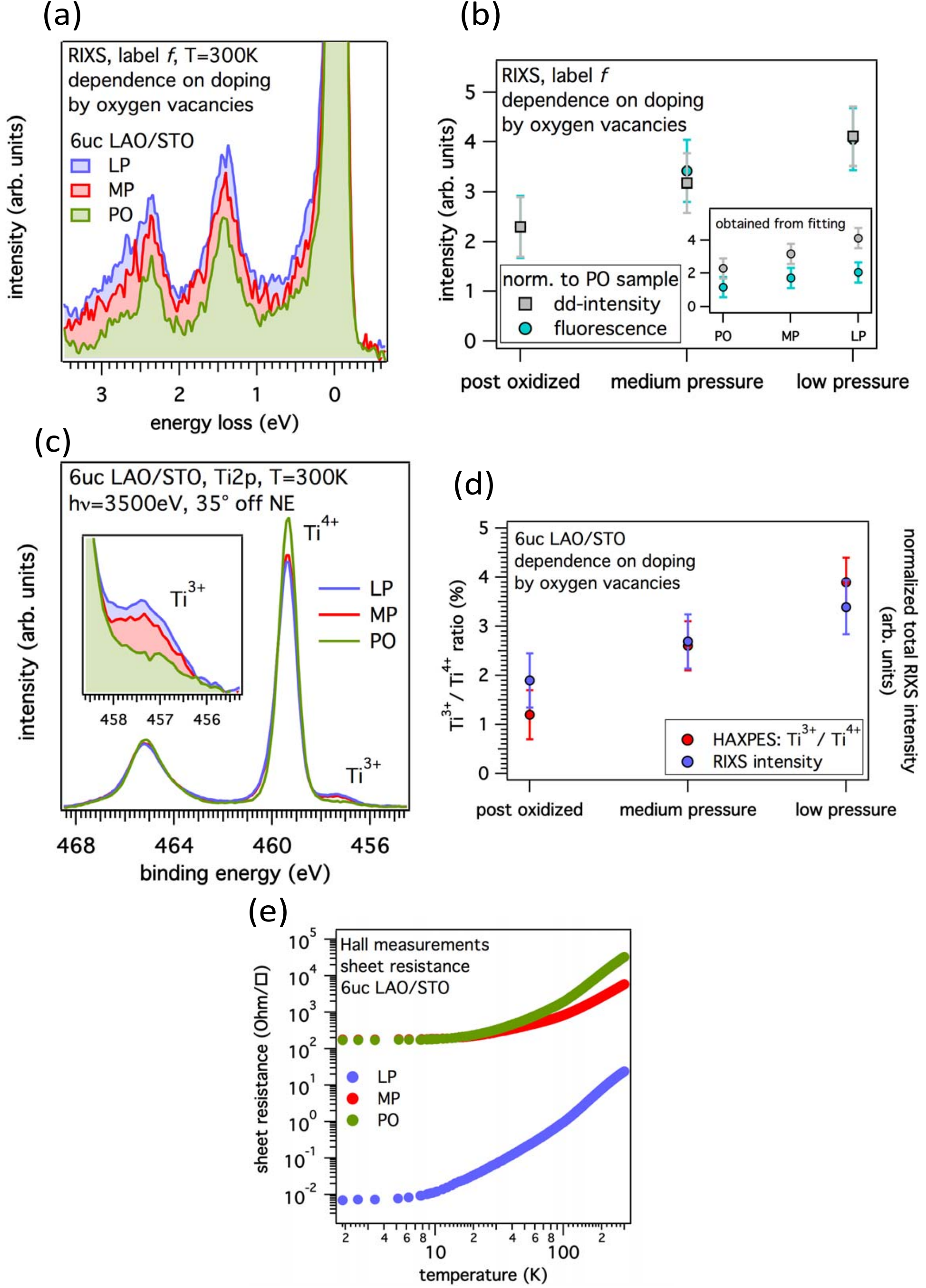}
	\caption{(a) Series of RIXS spectra for samples with varying oxygen 
vacancy concentrations normalized to the 3$d$$^1$$\underline{L}$ emission. PO, 
MP, and LP refer to post-growth oxygen annealing, growth in 
medium and in low oxygen pressure (see section \ref{sec:level2}). (b) 
Intensities 
of the $dd$-excitations and fluorescence-like signals normalized to the 
intensity of the PO sample obtained by fitting of the RIXS spectra. Both 
increase in parallel with the oxygen vacancy concentration. Intensities without 
normalization are shown in the inset. (c) HAXPES Ti~2$p$ spectra for the PO, MP, 
and LP samples at an emission angle of 35$^\circ$ off normal emission. (d) 
Comparison of the Ti$^{3+}$/Ti$^{4+}$ ratio obtained from fitting the Ti~2$p$ 
spectra and the total RIXS intensity (sum of $dd$-excitations $and$ 
fluorescence-like feature). RIXS data points have been normalized by minimizing 
the root-mean square deviation to the Ti$^{3+}$/Ti$^{4+}$ ratio of the three 
samples. (e) Temperature dependent sheet resistance of the PO, MP, and LP 
samples.}
	\label{fig4}
\end{figure}

Since oxygen vacancies act as electron donors we also studied 6\,uc samples 
with an intentionally higher amount of oxygen vacancies to further clarify the 
origin and character of the two observed RIXS excitations. In 
Fig.~\ref{fig4}~(a) the RIXS data --- normalized to 3$d$$^1$\underline{$L$} 
charge transfer excitations above 3.5\,eV (not shown) --- show the increase of 
the RIXS spectral weight as a function of oxygen vacancy concentration. As 
obtained from a fitting procedure equivalent to the one used for the thickness 
dependent series of spectra in Fig.~\ref{fig3}~(c), the fluorescence-like 
component and the integrated spectral weight of the $dd$-excitations 
rise in the same proportion with the amount of oxygen vacancies [see 
Fig.~\ref{fig4}~(b)]. Since both inelastic features are present in the 
spectrum for the fully oxidized sample (PO) neither component can be directly 
correlated to the amount of oxygen vacancies.

Figure~\ref{fig4}~(c) shows corresponding HAXPES Ti~2$p$ core-level 
spectra normalized to the overall integrated intensity. Beside a strong 
Ti~2$p$$_{3/2}$ peak due to emission from Ti$^{4+}$ ions mainly in the STO 
bulk, a chemically shifted contribution from Ti$^{3+}$-ions at the interface is 
visible at smaller binding energies. This signal is a measure for the 
extra electrons occupying the otherwise empty Ti~3$d$ states and therefore 
reflects the total amount of charge carriers in the system.\cite{Sing2009} By 
increasing the amount of oxygen vacancies a strong increase in the total amount 
of charge carriers is observed in HAXPES represented by the increase of the 
Ti$^{3+}$ contribution in the Ti~2$p$ spectrum as shown in the inset of 
Fig.~\ref{fig4}~(c).

Figure~\ref{fig4}~(d) displays the total RIXS intensity (sum of the integrated 
weights of the $dd$-excitations and fluorescence-like features) and the 
Ti$^{3+}$/Ti$^{4+}$ ratio obtained from the fitting of the Ti~2$p$ core-level 
HAXPES spectra of the three samples. After normalizing the RIXS to the HAXPES 
data by minimizing the root-mean square deviation of the three data points 
similar to Ref.~\onlinecite{Berner2010}, both spectroscopic techniques display 
qualitative agreement within experimental accuracy. These results suggest 
that the $total$ RIXS intensity also reflects the $total$ sheet carrier density 
as the Ti$^{3+}$ contribution in HAXPES.

According to Ref.~\onlinecite{Lin2013}, an oxygen vacancy in STO induces a 
localized level of mainly Ti~3$d$ character that can trap one electron, while 
the second electron becomes itinerant since double occupancy is forbidden by the 
strong onsite Coulomb interaction. As both RIXS components, $dd$ excitations 
and fluorescence, increase in parallel upon doping [see Fig.~\ref{fig4}~(b)] 
one could be misled to assume that one signal originates from localized charge 
carriers ($dd$-excitations), while the other feature reflects delocalized 
electrons (fluorescence) which would fit to this 1:1 doping rule. However, we 
have already ruled out this 
scenario for the reasons explained above.

As can be seen from resistivity measurements [see Fig.~\ref{fig4}~(e)] the 
effect of oxygen vacancies on the transport properties seems to be considerably 
smaller when skipping the post growth oxidation procedure (compare PO to MP 
data) than the effect of reducing the oxygen growth pressure (compare MP to LP 
data) for our samples. Since the charge carrier mobility at room temperature 
is essentially constant, it is thus a good measure for the charge carrier 
concentration that is determined to be 4.6\,x\,10$^{13}$\,cm$^{-2}$ for the PO, 
2.4\,x\,10$^{14}$\,cm$^{-2}$ for the MP, and 6.25\,x\,10$^{16}$\,cm$^{-2}$ for 
the LP sample. Note that a strong increase of the 
mobile charge carrier concentration is observed for the LP sample since the 
conducting region is not confined to the interface anymore but comprises the 
bulk substrate. If we stick to the 1:1 doping rule and assume that we can 
attribute the two types of RIXS features, $dd$ excitations and fluorescence, to 
localized and mobile charge carriers in the initial state, respectively, the 
intensity increase of both RIXS components should be significantly stronger 
between the MP and LP samples than between the PO and MP samples as expected 
from the results in Fig.~\ref{fig4}~(e). However, this expectation is not 
confirmed by our RIXS data in Figs.~\ref{fig4}~(a) and (b).

Furthermore, the evolution of the $total$ Ti~3$d$ charge carrier 
concentration obtained from HAXPES and the trend for the mobile charge carriers 
from resistivity measurements actually suggest that for larger oxygen vacancy 
concentrations the 1:1 doping rule is no longer valid. This deviation could be 
explained by the clustering of oxygen vacancies which alters the ratio of 
localized and delocalized charge carriers as suggested by density functional 
theory calculations.\cite{Jeschke2015} and should result in a change of the 
ratio of Raman- $and$ fluorescence-like signal. Again this is not seen in our 
RIXS data [Figs.~\ref{fig4}~(a) and (b)] requires a new interpretation.

\section{\label{sec:level4}Discussion}
The different behavior of the two components in RIXS of LAO/STO 
heterostructures upon varying the photon energy across the Ti-$L$ edge point 
at a 
different origin of these features. One signal can be ascribed to a 
Raman process giving rise to a constant energy loss, while the other feature 
shifts linearly with excitation energy and hence is due to a fluorescence-like 
channel. A previous assignment of these two features to localized and itinerant 
Ti~3$d$ charge carriers is clearly disproven by our present data. The 
observation of both RIXS components for the insulating 3\,uc sample and the 
saturation of both features at an overlayer thickness of 6\,uc as well as the 
inconsistent increase of the Raman- $and$ fluorescence-like intensities compared 
with the results of transport and HAXPES data upon oxygen vacancy doping 
call for a different interpretation.

\begin{figure}[t]
	\includegraphics[width=0.49\textwidth]{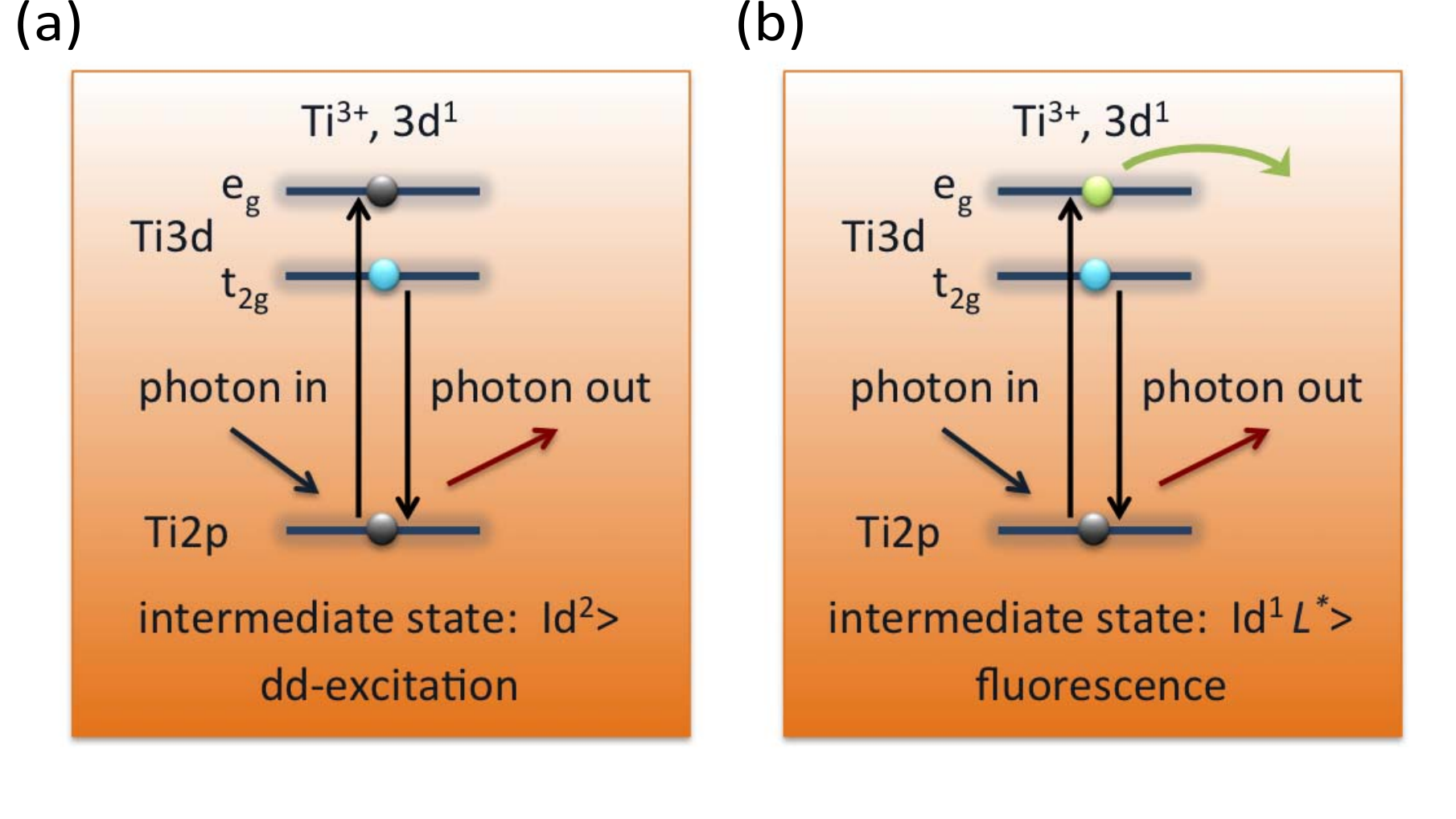}
	\caption{Schematic drawing of two possible RIXS processes with 
different intermediate states. (a) The excited electron stays localized in the 
$e_g$ states. After the decay process the final state is 3$d$$^1_{e_g}$ 
($dd$-excitation, Raman-signal). (b) The excited electron gets immediately 
delocalized which results in a situation similar to x-ray fluorescence. Here, 
the final state is 3$d$$^0$ (fluorescence-like-signal). Our results can be 
explained by a superposition of these two different channels in the intermediate 
state, 
$\alpha$$\vert$2$p$$^5$3$d$$^2$$>$+$\beta$$\vert$2$p$$^5$3$d$$^1$$L$$^*$$>$.}
	\label{fig5}
\end{figure}

To form such an alternative view it is helpful to analyse the nature of 
the possible intermediate states in the RIXS experiment after the initial 
photoexcitation of a core electron into the Ti~3$d$ shell (see Fig.~\ref{fig5}). 
Two type of intermediate states may result. Firstly, the intermediate state 
can be of the type $2p^53d^1t_{2g}3d^1e_g$ [see Fig.~\ref{fig5}~(a)]. In 
this case, both Ti~3$d$ electrons stay localized at the same atomic site. By 
deexcitation of one 3$d$ electron, the core hole created in the intial 
absorption is filled up again under emission of a photon. The energy difference 
between incoming and emitted photon reflects the associated, 
electronic $dd$-excitation between the 3$d$ $t_{2g}$- and the $e_g$-orbitals 
and appears in the RIXS spectra as a peak at finite loss energy with respect to 
the elastic line. In this case, the energy loss is independent of the incoming 
photon energy.

Secondly, assuming that one Ti~3$d$ electron gets delocalized following the 
photoexcitation, the intermediate state is of the type $2p^53d^1t_{2g}L^*$ with 
$L$$^*$ denoting an orbital of a neighboring ligand atom with an extra electron 
[see Fig.~\ref{fig5}~(b)]. The probability for this intermediate state is the 
higher, the more 
strongly the O~2$p$ ligand orbitals are hybridized with the Ti~3$d$ 
states. After the radiative decay, the final state is of the type $2p^63d^0$ 
and corresponds to x-ray fluorescence. Here, a 3$d$ valence electron decays 
radiatively into the $2p$ core hole generated by the preceding photoexcitation 
of an electron into continuum states above the Fermi level. The energy of 
the fluorescent x-ray photon in such a transition is independent of the 
excitation energy and the intensity reflects essentially the occupied 
$3d^1t_{2g}$ density of states.

Taking both decay channels into account, the intermediate state can be 
generally written as 
$\alpha$$\vert$2$p$$^5$3$d$$^2$$>$+$\beta$$\vert$2$p$$^5$3$d$$^1$$L$$^*$$>$, 
where the relative weight of the Raman and fluorescence-like signal is given by 
$\vert$$\alpha$$\vert$$^2$ and $\vert$$\beta$$\vert$$^2$ (with 
$\vert$$\alpha$$\vert$$^2$+$\vert$$\beta$$\vert$$^2$=1), respectively, and 
essentially determined by the hybridization strength of O~2$p$ and Ti~3$d$ 
orbitals. Therefore, the observation of either the Raman- or the 
fluorescence-like signal reflects the electronic character of the intermediate 
states instead of the itinerancy or localized character of the Ti\,3$d$ 
electrons in the ground state.

Our observation and explanation of the two RIXS features due to $dd$ 
excitations and fluorescence are in full agreement with the experimental RIXS 
studies on BaTiO$_3$ and BaSO$_4$\cite{Yoshii2012} as well as 
theoretical studies of the cluster size effect on resonant x-ray emission 
spectra by Ide and Kotani.\cite{Ide1998,Ide2000} If the cluster-size is small, 
the excited electron is essentially localized. In larger clusters, 
intermediate states can occur in which the excited conduction electron is 
extended in space.

\section{\label{sec:level5}Summary}
We have investigated LAO/STO heterostructures with varying LAO overlayer 
thickness and oxygen vacancy concentration by means of RIXS at the Ti-$L$ edge. 
Already for the sample with a 3\,uc thick film, i.e., below the critical 
thickness for conductivity, we have observed finite 
spectral weight, signalling that localized charge 
carriers are present in the ground state. This finding is in line with the 
report on ferromagnetism which has been ascribed to the local moments of 
trapped electrons. As seen previously, we also have observed two 
types of contributions to the RIXS spectra, namely due to $dd$ excitations and 
fluorescence. They behave differently upon the variation of the 
excitation energy. A detailed analysis for samples with different 
film thicknesses and oxygen vacancy concentrations in combination with the 
results of HAXPES and transport experiments, reveals the physical meaning of 
these two contributions in contrast to previous claims. The appearance of 
fluorescence in addition to Raman features reflects the probability for the 
excited electron to escape from the Ti ion in the coherent RIXS process. Thus 
the RIXS spectra provide information on the intermediate state rather than the 
ground state properties. Furthermore, by 
comparing the $total$ RIXS intensity (sum of $dd$-excitations and 
fluorescence-like signal) and the Ti$^{3+}$/Ti$^{4+}$ ratio obtained from HAXPES 
measurements, we conclude that the $total$ RIXS intensity reflects the $total$ 
amount of charge carriers (localized $and$ delocalized) present in the system 
within experimental accuracy.

\section{\label{sec:level6}Acknowledgments}
Funding by the Deutsche Forschungsgemeinschaft (FOR 1162) is gratefully 
acknowledged. Experiments at SPring-8 BL07LSU were performed jointly by the 
Synchrotron 
Radiation Research Organization and the University of Tokyo (Proposal No. 
2012A7429, 2012B7439, 2013A7449). This work was financially supported by a 
Grants-in-Aid for Young Scientists (23740240) from the Ministry of Education, 
Culture, Sports, Science and Technology, Japan.  We thank Y. Nishitani and Y. 
Nakata for supporting the experiments.


\begin{thebibliography}{37}%
\makeatletter
\providecommand \@ifxundefined [1]{%
 \@ifx{#1\undefined}
}%
\providecommand \@ifnum [1]{%
 \ifnum #1\expandafter \@firstoftwo
 \else \expandafter \@secondoftwo
 \fi
}%
\providecommand \@ifx [1]{%
 \ifx #1\expandafter \@firstoftwo
 \else \expandafter \@secondoftwo
 \fi
}%
\providecommand \natexlab [1]{#1}%
\providecommand \enquote  [1]{``#1''}%
\providecommand \bibnamefont  [1]{#1}%
\providecommand \bibfnamefont [1]{#1}%
\providecommand \citenamefont [1]{#1}%
\providecommand \href@noop [0]{\@secondoftwo}%
\providecommand \href [0]{\begingroup \@sanitize@url \@href}%
\providecommand \@href[1]{\@@startlink{#1}\@@href}%
\providecommand \@@href[1]{\endgroup#1\@@endlink}%
\providecommand \@sanitize@url [0]{\catcode `\\12\catcode `\$12\catcode
  `\&12\catcode `\#12\catcode `\^12\catcode `\_12\catcode `\%12\relax}%
\providecommand \@@startlink[1]{}%
\providecommand \@@endlink[0]{}%
\providecommand \url  [0]{\begingroup\@sanitize@url \@url }%
\providecommand \@url [1]{\endgroup\@href {#1}{\urlprefix }}%
\providecommand \urlprefix  [0]{URL }%
\providecommand \Eprint [0]{\href }%
\providecommand \doibase [0]{http://dx.doi.org/}%
\providecommand \selectlanguage [0]{\@gobble}%
\providecommand \bibinfo  [0]{\@secondoftwo}%
\providecommand \bibfield  [0]{\@secondoftwo}%
\providecommand \translation [1]{[#1]}%
\providecommand \BibitemOpen [0]{}%
\providecommand \bibitemStop [0]{}%
\providecommand \bibitemNoStop [0]{.\EOS\space}%
\providecommand \EOS [0]{\spacefactor3000\relax}%
\providecommand \BibitemShut  [1]{\csname bibitem#1\endcsname}%
\let\auto@bib@innerbib\@empty
%</preamble>
\bibitem [{\citenamefont {Othomo}\ and\ \citenamefont
  {Hwang}(2004)}]{Ohtomo2002}%
  \BibitemOpen
  \bibfield  {author} {\bibinfo {author} {\bibfnamefont {A.}~\bibnamefont
  {Othomo}}\ and\ \bibinfo {author} {\bibfnamefont {H.~Y.}\ \bibnamefont
  {Hwang}},\ }\href@noop {} {\bibfield  {journal} {\bibinfo  {journal}
  {Nature}\ }\textbf {\bibinfo {volume} {427}},\ \bibinfo {pages} {423}
  (\bibinfo {year} {2004})}\BibitemShut {NoStop}%
\bibitem [{\citenamefont {Thiel}\ \emph {et~al.}(2006)\citenamefont {Thiel},
  \citenamefont {Hammerl}, \citenamefont {Schmehl}, \citenamefont {Schneider},\
  and\ \citenamefont {Mannhart}}]{Thiel2006}%
  \BibitemOpen
  \bibfield  {author} {\bibinfo {author} {\bibfnamefont {S.}~\bibnamefont
  {Thiel}}, \bibinfo {author} {\bibfnamefont {G.}~\bibnamefont {Hammerl}},
  \bibinfo {author} {\bibfnamefont {A.}~\bibnamefont {Schmehl}}, \bibinfo
  {author} {\bibfnamefont {C.~W.}\ \bibnamefont {Schneider}}, \ and\ \bibinfo
  {author} {\bibfnamefont {J.}~\bibnamefont {Mannhart}},\ }\href@noop {}
  {\bibfield  {journal} {\bibinfo  {journal} {Science}\ }\textbf {\bibinfo
  {volume} {313}},\ \bibinfo {pages} {1942} (\bibinfo {year}
  {2006})}\BibitemShut {NoStop}%
\bibitem [{\citenamefont {Sing}\ \emph {et~al.}(2009)\citenamefont {Sing},
  \citenamefont {Berner}, \citenamefont {Go{\ss}}, \citenamefont
  {A.M{\"u}ller}, \citenamefont {Ruff}, \citenamefont {Wetscherek},
  \citenamefont {Thiel}, \citenamefont {Mannhart}, \citenamefont {Pauli},
  \citenamefont {Schneider}, \citenamefont {Willmott}, \citenamefont {Gorgoi},
  \citenamefont {Sch{\"a}fers},\ and\ \citenamefont {Claessen}}]{Sing2009}%
  \BibitemOpen
  \bibfield  {author} {\bibinfo {author} {\bibfnamefont {M.}~\bibnamefont
  {Sing}}, \bibinfo {author} {\bibfnamefont {G.}~\bibnamefont {Berner}},
  \bibinfo {author} {\bibfnamefont {K.}~\bibnamefont {Go{\ss}}}, \bibinfo
  {author} {\bibnamefont {A.M{\"u}ller}}, \bibinfo {author} {\bibfnamefont
  {A.}~\bibnamefont {Ruff}}, \bibinfo {author} {\bibfnamefont {A.}~\bibnamefont
  {Wetscherek}}, \bibinfo {author} {\bibfnamefont {S.}~\bibnamefont {Thiel}},
  \bibinfo {author} {\bibfnamefont {J.}~\bibnamefont {Mannhart}}, \bibinfo
  {author} {\bibfnamefont {S.~A.}\ \bibnamefont {Pauli}}, \bibinfo {author}
  {\bibfnamefont {C.~W.}\ \bibnamefont {Schneider}}, \bibinfo {author}
  {\bibfnamefont {P.~R.}\ \bibnamefont {Willmott}}, \bibinfo {author}
  {\bibfnamefont {M.}~\bibnamefont {Gorgoi}}, \bibinfo {author} {\bibfnamefont
  {F.}~\bibnamefont {Sch{\"a}fers}}, \ and\ \bibinfo {author} {\bibfnamefont
  {R.}~\bibnamefont {Claessen}},\ }\href@noop {} {\bibfield  {journal}
  {\bibinfo  {journal} {Phys. Rev. Lett.}\ }\textbf {\bibinfo {volume} {102}},\
  \bibinfo {pages} {176805} (\bibinfo {year} {2009})}\BibitemShut {NoStop}%
\bibitem [{\citenamefont {Caviglia}\ \emph {et~al.}(2008)\citenamefont
  {Caviglia}, \citenamefont {Gariglio}, \citenamefont {Reyren}, \citenamefont
  {Jaccard}, \citenamefont {Schneider}, \citenamefont {Gabay}, \citenamefont
  {Thiel}, \citenamefont {Hammerl}, \citenamefont {Mannhart},\ and\
  \citenamefont {Triscone}}]{Caviglia2008}%
  \BibitemOpen
  \bibfield  {author} {\bibinfo {author} {\bibfnamefont {A.~D.}\ \bibnamefont
  {Caviglia}}, \bibinfo {author} {\bibfnamefont {S.}~\bibnamefont {Gariglio}},
  \bibinfo {author} {\bibfnamefont {N.}~\bibnamefont {Reyren}}, \bibinfo
  {author} {\bibfnamefont {D.}~\bibnamefont {Jaccard}}, \bibinfo {author}
  {\bibfnamefont {T.}~\bibnamefont {Schneider}}, \bibinfo {author}
  {\bibfnamefont {M.}~\bibnamefont {Gabay}}, \bibinfo {author} {\bibfnamefont
  {S.}~\bibnamefont {Thiel}}, \bibinfo {author} {\bibfnamefont
  {G.}~\bibnamefont {Hammerl}}, \bibinfo {author} {\bibfnamefont
  {J.}~\bibnamefont {Mannhart}}, \ and\ \bibinfo {author} {\bibfnamefont
  {J.-M.}\ \bibnamefont {Triscone}},\ }\href@noop {} {\bibfield  {journal}
  {\bibinfo  {journal} {Nature}\ }\textbf {\bibinfo {volume} {456}},\ \bibinfo
  {pages} {624} (\bibinfo {year} {2008})}\BibitemShut {NoStop}%
\bibitem [{\citenamefont {Reyren}\ \emph {et~al.}(2007)\citenamefont {Reyren},
  \citenamefont {Thiel}, \citenamefont {Caviglia}, \citenamefont {Kourkoutis},
  \citenamefont {Hammerl}, \citenamefont {Richter}, \citenamefont {Schneider},
  \citenamefont {Kopp}, \citenamefont {R{\"u}etschi}, \citenamefont {Jaccard},
  \citenamefont {Gabay}, \citenamefont {Muller}, \citenamefont {Triscone},\
  and\ \citenamefont {Mannhart}}]{Reyren2007}%
  \BibitemOpen
  \bibfield  {author} {\bibinfo {author} {\bibfnamefont {N.}~\bibnamefont
  {Reyren}}, \bibinfo {author} {\bibfnamefont {S.}~\bibnamefont {Thiel}},
  \bibinfo {author} {\bibfnamefont {A.~D.}\ \bibnamefont {Caviglia}}, \bibinfo
  {author} {\bibfnamefont {L.~F.}\ \bibnamefont {Kourkoutis}}, \bibinfo
  {author} {\bibfnamefont {G.}~\bibnamefont {Hammerl}}, \bibinfo {author}
  {\bibfnamefont {C.}~\bibnamefont {Richter}}, \bibinfo {author} {\bibfnamefont
  {C.~W.}\ \bibnamefont {Schneider}}, \bibinfo {author} {\bibfnamefont
  {T.}~\bibnamefont {Kopp}}, \bibinfo {author} {\bibfnamefont {A.-S.}\
  \bibnamefont {R{\"u}etschi}}, \bibinfo {author} {\bibfnamefont
  {D.}~\bibnamefont {Jaccard}}, \bibinfo {author} {\bibfnamefont
  {M.}~\bibnamefont {Gabay}}, \bibinfo {author} {\bibfnamefont {D.~A.}\
  \bibnamefont {Muller}}, \bibinfo {author} {\bibfnamefont {J.-M.}\
  \bibnamefont {Triscone}}, \ and\ \bibinfo {author} {\bibfnamefont
  {J.}~\bibnamefont {Mannhart}},\ }\href@noop {} {\bibfield  {journal}
  {\bibinfo  {journal} {Science}\ }\textbf {\bibinfo {volume} {317}},\ \bibinfo
  {pages} {1196} (\bibinfo {year} {2007})}\BibitemShut {NoStop}%
\bibitem [{\citenamefont {Li}\ \emph {et~al.}(2011)\citenamefont {Li},
  \citenamefont {Richter}, \citenamefont {Mannhart},\ and\ \citenamefont
  {Ashoori}}]{Li2011}%
  \BibitemOpen
  \bibfield  {author} {\bibinfo {author} {\bibfnamefont {L.}~\bibnamefont
  {Li}}, \bibinfo {author} {\bibfnamefont {C.}~\bibnamefont {Richter}},
  \bibinfo {author} {\bibfnamefont {J.}~\bibnamefont {Mannhart}}, \ and\
  \bibinfo {author} {\bibfnamefont {R.~C.}\ \bibnamefont {Ashoori}},\
  }\href@noop {} {\bibfield  {journal} {\bibinfo  {journal} {Nature Physics}\
  }\textbf {\bibinfo {volume} {7}},\ \bibinfo {pages} {762} (\bibinfo {year}
  {2011})}\BibitemShut {NoStop}%
\bibitem [{\citenamefont {Bert}\ \emph {et~al.}(2011)\citenamefont {Bert},
  \citenamefont {Kalisky}, \citenamefont {Bell}, \citenamefont {Kim},
  \citenamefont {Hikita}, \citenamefont {Hwang},\ and\ \citenamefont
  {Moler}}]{Bert2011}%
  \BibitemOpen
  \bibfield  {author} {\bibinfo {author} {\bibfnamefont {J.~A.}\ \bibnamefont
  {Bert}}, \bibinfo {author} {\bibfnamefont {B.}~\bibnamefont {Kalisky}},
  \bibinfo {author} {\bibfnamefont {C.}~\bibnamefont {Bell}}, \bibinfo {author}
  {\bibfnamefont {M.}~\bibnamefont {Kim}}, \bibinfo {author} {\bibfnamefont
  {Y.}~\bibnamefont {Hikita}}, \bibinfo {author} {\bibfnamefont {H.~Y.}\
  \bibnamefont {Hwang}}, \ and\ \bibinfo {author} {\bibfnamefont {K.~A.}\
  \bibnamefont {Moler}},\ }\href@noop {} {\bibfield  {journal} {\bibinfo
  {journal} {Nat. Phys.}\ }\textbf {\bibinfo {volume} {7}},\ \bibinfo {pages}
  {767} (\bibinfo {year} {2011})}\BibitemShut {NoStop}%
\bibitem [{\citenamefont {Kalisky}\ \emph {et~al.}(2012)\citenamefont
  {Kalisky}, \citenamefont {Bert}, \citenamefont {Klopfer}, \citenamefont
  {Bell}, \citenamefont {Sato}, \citenamefont {Hosoda}, \citenamefont {Hikita},
  \citenamefont {Hwang},\ and\ \citenamefont {Moler}}]{Kalisky2012}%
  \BibitemOpen
  \bibfield  {author} {\bibinfo {author} {\bibfnamefont {B.}~\bibnamefont
  {Kalisky}}, \bibinfo {author} {\bibfnamefont {J.~A.}\ \bibnamefont {Bert}},
  \bibinfo {author} {\bibfnamefont {B.~B.}\ \bibnamefont {Klopfer}}, \bibinfo
  {author} {\bibfnamefont {C.}~\bibnamefont {Bell}}, \bibinfo {author}
  {\bibfnamefont {H.~K.}\ \bibnamefont {Sato}}, \bibinfo {author}
  {\bibfnamefont {M.}~\bibnamefont {Hosoda}}, \bibinfo {author} {\bibfnamefont
  {Y.}~\bibnamefont {Hikita}}, \bibinfo {author} {\bibfnamefont {H.~Y.}\
  \bibnamefont {Hwang}}, \ and\ \bibinfo {author} {\bibfnamefont {K.~A.}\
  \bibnamefont {Moler}},\ }\href@noop {} {\bibfield  {journal} {\bibinfo
  {journal} {Nat. commun.}\ }\textbf {\bibinfo {volume} {3}},\ \bibinfo {pages}
  {922} (\bibinfo {year} {2012})}\BibitemShut {NoStop}%
\bibitem [{\citenamefont {Popovic}\ \emph {et~al.}(2008)\citenamefont
  {Popovic}, \citenamefont {Satpathy},\ and\ \citenamefont
  {Martin}}]{Popovic2008}%
  \BibitemOpen
  \bibfield  {author} {\bibinfo {author} {\bibfnamefont {Z.~S.}\ \bibnamefont
  {Popovic}}, \bibinfo {author} {\bibfnamefont {S.}~\bibnamefont {Satpathy}}, \
  and\ \bibinfo {author} {\bibfnamefont {R.~M.}\ \bibnamefont {Martin}},\
  }\href@noop {} {\bibfield  {journal} {\bibinfo  {journal} {Phys. Rev. Lett.}\
  }\textbf {\bibinfo {volume} {101}},\ \bibinfo {pages} {256801} (\bibinfo
  {year} {2008})}\BibitemShut {NoStop}%
\bibitem [{\citenamefont {Chan}\ \emph {et~al.}(2009)\citenamefont {Chan},
  \citenamefont {Werner},\ and\ \citenamefont {Millis}}]{Chan2009}%
  \BibitemOpen
  \bibfield  {author} {\bibinfo {author} {\bibfnamefont {C.-K.}\ \bibnamefont
  {Chan}}, \bibinfo {author} {\bibfnamefont {P.}~\bibnamefont {Werner}}, \ and\
  \bibinfo {author} {\bibfnamefont {A.~J.}\ \bibnamefont {Millis}},\
  }\href@noop {} {\bibfield  {journal} {\bibinfo  {journal} {Phys. Rev. B.}\
  }\textbf {\bibinfo {volume} {80}},\ \bibinfo {pages} {235114} (\bibinfo
  {year} {2009})}\BibitemShut {NoStop}%
\bibitem [{\citenamefont {Elfimov}\ \emph {et~al.}(2002)\citenamefont
  {Elfimov}, \citenamefont {Yunoki},\ and\ \citenamefont
  {Sawatzky}}]{Elfimov2002}%
  \BibitemOpen
  \bibfield  {author} {\bibinfo {author} {\bibfnamefont {I.~S.}\ \bibnamefont
  {Elfimov}}, \bibinfo {author} {\bibfnamefont {S.}~\bibnamefont {Yunoki}}, \
  and\ \bibinfo {author} {\bibfnamefont {G.~A.}\ \bibnamefont {Sawatzky}},\
  }\href@noop {} {\bibfield  {journal} {\bibinfo  {journal} {Phys. Rev. Lett.}\
  }\textbf {\bibinfo {volume} {89}},\ \bibinfo {pages} {216403} (\bibinfo
  {year} {2002})}\BibitemShut {NoStop}%
\bibitem [{\citenamefont {Pavlenko}\ \emph {et~al.}(2012)\citenamefont
  {Pavlenko}, \citenamefont {Kopp}, \citenamefont {Tsymbal}, \citenamefont
  {Sawatzky},\ and\ \citenamefont {Mannhart}}]{Pavlenko2012}%
  \BibitemOpen
  \bibfield  {author} {\bibinfo {author} {\bibfnamefont {N.}~\bibnamefont
  {Pavlenko}}, \bibinfo {author} {\bibfnamefont {T.}~\bibnamefont {Kopp}},
  \bibinfo {author} {\bibfnamefont {E.~Y.}\ \bibnamefont {Tsymbal}}, \bibinfo
  {author} {\bibfnamefont {G.~A.}\ \bibnamefont {Sawatzky}}, \ and\ \bibinfo
  {author} {\bibfnamefont {J.}~\bibnamefont {Mannhart}},\ }\href@noop {}
  {\bibfield  {journal} {\bibinfo  {journal} {Phys. Rev. B.}\ }\textbf
  {\bibinfo {volume} {85}},\ \bibinfo {pages} {020407 (R)} (\bibinfo {year}
  {2012})}\BibitemShut {NoStop}%
\bibitem [{\citenamefont {Nakagawa}\ \emph {et~al.}(2006)\citenamefont
  {Nakagawa}, \citenamefont {Hwang},\ and\ \citenamefont
  {Muller}}]{Nakagawa2006}%
  \BibitemOpen
  \bibfield  {author} {\bibinfo {author} {\bibfnamefont {N.}~\bibnamefont
  {Nakagawa}}, \bibinfo {author} {\bibfnamefont {H.~Y.}\ \bibnamefont {Hwang}},
  \ and\ \bibinfo {author} {\bibfnamefont {D.~A.}\ \bibnamefont {Muller}},\
  }\href@noop {} {\bibfield  {journal} {\bibinfo  {journal} {Nature Mater.}\
  }\textbf {\bibinfo {volume} {5}},\ \bibinfo {pages} {204} (\bibinfo {year}
  {2006})}\BibitemShut {NoStop}%
\bibitem [{\citenamefont {Yu}\ and\ \citenamefont {Zunger}(2014)}]{yu2014}%
  \BibitemOpen
  \bibfield  {author} {\bibinfo {author} {\bibfnamefont {L.}~\bibnamefont
  {Yu}}\ and\ \bibinfo {author} {\bibfnamefont {A.}~\bibnamefont {Zunger}},\
  }\href@noop {} {\bibfield  {journal} {\bibinfo  {journal} {Nat. Commun.}\
  }\textbf {\bibinfo {volume} {5}},\ \bibinfo {pages} {5118} (\bibinfo {year}
  {2014})}\BibitemShut {NoStop}%
\bibitem [{\citenamefont {Kalabukhov}\ \emph {et~al.}(2007)\citenamefont
  {Kalabukhov}, \citenamefont {Gunnarsson}, \citenamefont {Börjesson},
  \citenamefont {Olsson}, \citenamefont {Claeson},\ and\ \citenamefont
  {Winkler}}]{kalabukhov2007}%
  \BibitemOpen
  \bibfield  {author} {\bibinfo {author} {\bibfnamefont {A.}~\bibnamefont
  {Kalabukhov}}, \bibinfo {author} {\bibfnamefont {R.}~\bibnamefont
  {Gunnarsson}}, \bibinfo {author} {\bibfnamefont {J.}~\bibnamefont
  {Börjesson}}, \bibinfo {author} {\bibfnamefont {E.}~\bibnamefont {Olsson}},
  \bibinfo {author} {\bibfnamefont {T.}~\bibnamefont {Claeson}}, \ and\
  \bibinfo {author} {\bibfnamefont {D.}~\bibnamefont {Winkler}},\ }\href
  {http://link.aps.org/doi/10.1103/PhysRevB.75.121404} {\bibfield  {journal}
  {\bibinfo  {journal} {Phys. Rev. B}\ }\textbf {\bibinfo {volume} {75}}
  (\bibinfo {year} {2007})}\BibitemShut {NoStop}%
\bibitem [{\citenamefont {Liu}\ \emph {et~al.}(2013)\citenamefont {Liu},
  \citenamefont {Li}, \citenamefont {L\"u}, \citenamefont {Huang},
  \citenamefont {Huang}, \citenamefont {Zeng}, \citenamefont {Qiu},
  \citenamefont {Huang}, \citenamefont {Annadi}, \citenamefont {Chen},
  \citenamefont {Coey}, \citenamefont {Venkatesan},\ and\ \citenamefont
  {{Ariando}}}]{liu2013}%
  \BibitemOpen
  \bibfield  {author} {\bibinfo {author} {\bibfnamefont {Z.~Q.}\ \bibnamefont
  {Liu}}, \bibinfo {author} {\bibfnamefont {C.~J.}\ \bibnamefont {Li}},
  \bibinfo {author} {\bibfnamefont {W.~M.}\ \bibnamefont {L\"u}}, \bibinfo
  {author} {\bibfnamefont {X.~H.}\ \bibnamefont {Huang}}, \bibinfo {author}
  {\bibfnamefont {Z.}~\bibnamefont {Huang}}, \bibinfo {author} {\bibfnamefont
  {S.~W.}\ \bibnamefont {Zeng}}, \bibinfo {author} {\bibfnamefont {X.~P.}\
  \bibnamefont {Qiu}}, \bibinfo {author} {\bibfnamefont {L.~S.}\ \bibnamefont
  {Huang}}, \bibinfo {author} {\bibfnamefont {A.}~\bibnamefont {Annadi}},
  \bibinfo {author} {\bibfnamefont {J.~S.}\ \bibnamefont {Chen}}, \bibinfo
  {author} {\bibfnamefont {J.~M.~D.}\ \bibnamefont {Coey}}, \bibinfo {author}
  {\bibfnamefont {T.}~\bibnamefont {Venkatesan}}, \ and\ \bibinfo {author}
  {\bibnamefont {{Ariando}}},\ }\href
  {http://link.aps.org/doi/10.1103/PhysRevX.3.021010} {\bibfield  {journal}
  {\bibinfo  {journal} {Phys. Rev. X}\ }\textbf {\bibinfo {volume} {3}}
  (\bibinfo {year} {2013})}\BibitemShut {NoStop}%
\bibitem [{\citenamefont {Cancellieri}\ \emph {et~al.}(2010)\citenamefont
  {Cancellieri}, \citenamefont {Reyren}, \citenamefont {Gariglio},
  \citenamefont {Caviglia}, \citenamefont {F\^{e}te},\ and\ \citenamefont
  {Triscone}}]{Cancellieri2010}%
  \BibitemOpen
  \bibfield  {author} {\bibinfo {author} {\bibfnamefont {C.}~\bibnamefont
  {Cancellieri}}, \bibinfo {author} {\bibfnamefont {N.}~\bibnamefont {Reyren}},
  \bibinfo {author} {\bibfnamefont {S.}~\bibnamefont {Gariglio}}, \bibinfo
  {author} {\bibfnamefont {A.~D.}\ \bibnamefont {Caviglia}}, \bibinfo {author}
  {\bibfnamefont {A.}~\bibnamefont {F\^{e}te}}, \ and\ \bibinfo {author}
  {\bibfnamefont {J.-M.}\ \bibnamefont {Triscone}},\ }\href@noop {} {\bibfield
  {journal} {\bibinfo  {journal} {Europhys. Lett.}\ }\textbf {\bibinfo {volume}
  {91}},\ \bibinfo {pages} {17004} (\bibinfo {year} {2010})}\BibitemShut
  {NoStop}%
\bibitem [{\citenamefont {Basletic}\ \emph {et~al.}(2008)\citenamefont
  {Basletic}, \citenamefont {Maurice}, \citenamefont {Carr\'et\'ero},
  \citenamefont {Herranz}, \citenamefont {Copie}, \citenamefont {Bibes},
  \citenamefont {Jacquet}, \citenamefont {Bouzehouane}, \citenamefont
  {s.~Fusil},\ and\ \citenamefont {Barth\'el\'emy}}]{Basletic2008}%
  \BibitemOpen
  \bibfield  {author} {\bibinfo {author} {\bibfnamefont {M.}~\bibnamefont
  {Basletic}}, \bibinfo {author} {\bibfnamefont {J.-L.}\ \bibnamefont
  {Maurice}}, \bibinfo {author} {\bibfnamefont {C.}~\bibnamefont
  {Carr\'et\'ero}}, \bibinfo {author} {\bibfnamefont {G.}~\bibnamefont
  {Herranz}}, \bibinfo {author} {\bibfnamefont {O.}~\bibnamefont {Copie}},
  \bibinfo {author} {\bibfnamefont {M.}~\bibnamefont {Bibes}}, \bibinfo
  {author} {\bibfnamefont {E.}~\bibnamefont {Jacquet}}, \bibinfo {author}
  {\bibfnamefont {K.}~\bibnamefont {Bouzehouane}}, \bibinfo {author}
  {\bibnamefont {s.~Fusil}}, \ and\ \bibinfo {author} {\bibfnamefont
  {A.}~\bibnamefont {Barth\'el\'emy}},\ }\href@noop {} {\bibfield  {journal}
  {\bibinfo  {journal} {Nature Mater.}\ }\textbf {\bibinfo {volume} {7}},\
  \bibinfo {pages} {621} (\bibinfo {year} {2008})}\BibitemShut {NoStop}%
\bibitem [{\citenamefont {Berner}\ \emph {et~al.}(2013)\citenamefont {Berner},
  \citenamefont {Sing}, \citenamefont {Fujiwara}, \citenamefont {Yasui},
  \citenamefont {Saitoh}, \citenamefont {Yamasaki}, \citenamefont {Nishitani},
  \citenamefont {Sekiyama}, \citenamefont {Pavlenko}, \citenamefont {Kopp},
  \citenamefont {Richter}, \citenamefont {Mannhart}, \citenamefont {Suga},\
  and\ \citenamefont {Claessen}}]{Berner2013}%
  \BibitemOpen
  \bibfield  {author} {\bibinfo {author} {\bibfnamefont {G.}~\bibnamefont
  {Berner}}, \bibinfo {author} {\bibfnamefont {M.}~\bibnamefont {Sing}},
  \bibinfo {author} {\bibfnamefont {H.}~\bibnamefont {Fujiwara}}, \bibinfo
  {author} {\bibfnamefont {A.}~\bibnamefont {Yasui}}, \bibinfo {author}
  {\bibfnamefont {Y.}~\bibnamefont {Saitoh}}, \bibinfo {author} {\bibfnamefont
  {A.}~\bibnamefont {Yamasaki}}, \bibinfo {author} {\bibfnamefont
  {Y.}~\bibnamefont {Nishitani}}, \bibinfo {author} {\bibfnamefont
  {A.}~\bibnamefont {Sekiyama}}, \bibinfo {author} {\bibfnamefont
  {N.}~\bibnamefont {Pavlenko}}, \bibinfo {author} {\bibfnamefont
  {T.}~\bibnamefont {Kopp}}, \bibinfo {author} {\bibfnamefont {C.}~\bibnamefont
  {Richter}}, \bibinfo {author} {\bibfnamefont {J.}~\bibnamefont {Mannhart}},
  \bibinfo {author} {\bibfnamefont {S.}~\bibnamefont {Suga}}, \ and\ \bibinfo
  {author} {\bibfnamefont {R.}~\bibnamefont {Claessen}},\ }\href@noop {}
  {\bibfield  {journal} {\bibinfo  {journal} {Phys. Rev. Lett.}\ }\textbf
  {\bibinfo {volume} {110}},\ \bibinfo {pages} {247601} (\bibinfo {year}
  {2013})}\BibitemShut {NoStop}%
\bibitem [{\citenamefont {Sch{\"u}tz}\ \emph {et~al.}(2015)\citenamefont
  {Sch{\"u}tz}, \citenamefont {Pfaff}, \citenamefont {Scheiderer},
  \citenamefont {Chen}, \citenamefont {Pryds}, \citenamefont {Gorgoi},
  \citenamefont {Sing},\ and\ \citenamefont {Claessen}}]{Schuetz22015}%
  \BibitemOpen
  \bibfield  {author} {\bibinfo {author} {\bibfnamefont {P.}~\bibnamefont
  {Sch{\"u}tz}}, \bibinfo {author} {\bibfnamefont {F.}~\bibnamefont {Pfaff}},
  \bibinfo {author} {\bibfnamefont {P.}~\bibnamefont {Scheiderer}}, \bibinfo
  {author} {\bibfnamefont {Y.~Z.}\ \bibnamefont {Chen}}, \bibinfo {author}
  {\bibfnamefont {N.}~\bibnamefont {Pryds}}, \bibinfo {author} {\bibfnamefont
  {M.}~\bibnamefont {Gorgoi}}, \bibinfo {author} {\bibfnamefont
  {M.}~\bibnamefont {Sing}}, \ and\ \bibinfo {author} {\bibfnamefont
  {R.}~\bibnamefont {Claessen}},\ }\href@noop {} {\bibfield  {journal}
  {\bibinfo  {journal} {Phys. Rev. B.}\ }\textbf {\bibinfo {volume} {91}},\
  \bibinfo {pages} {165118} (\bibinfo {year} {2015})}\BibitemShut {NoStop}%
\bibitem [{\citenamefont {Cancellieri}\ \emph {et~al.}(2014)\citenamefont
  {Cancellieri}, \citenamefont {Reinle-Schmitt}, \citenamefont {Kobayashi},
  \citenamefont {Strocov}, \citenamefont {Willmott}, \citenamefont {Fontaine},
  \citenamefont {Ghosez}, \citenamefont {Filippetti}, \citenamefont {Delugas},\
  and\ \citenamefont {Fiorentini}}]{Cancellieri2014}%
  \BibitemOpen
  \bibfield  {author} {\bibinfo {author} {\bibfnamefont {C.}~\bibnamefont
  {Cancellieri}}, \bibinfo {author} {\bibfnamefont {M.~L.}\ \bibnamefont
  {Reinle-Schmitt}}, \bibinfo {author} {\bibfnamefont {M.}~\bibnamefont
  {Kobayashi}}, \bibinfo {author} {\bibfnamefont {V.~N.}\ \bibnamefont
  {Strocov}}, \bibinfo {author} {\bibfnamefont {P.~R.}\ \bibnamefont
  {Willmott}}, \bibinfo {author} {\bibfnamefont {D.}~\bibnamefont {Fontaine}},
  \bibinfo {author} {\bibfnamefont {P.}~\bibnamefont {Ghosez}}, \bibinfo
  {author} {\bibfnamefont {A.}~\bibnamefont {Filippetti}}, \bibinfo {author}
  {\bibfnamefont {P.}~\bibnamefont {Delugas}}, \ and\ \bibinfo {author}
  {\bibfnamefont {V.}~\bibnamefont {Fiorentini}},\ }\href@noop {} {\bibfield
  {journal} {\bibinfo  {journal} {Phys. Rev. B.}\ }\textbf {\bibinfo {volume}
  {89}},\ \bibinfo {pages} {121412(R)} (\bibinfo {year} {2014})}\BibitemShut
  {NoStop}%
\bibitem [{\citenamefont {Berner}\ \emph {et~al.}(2010)\citenamefont {Berner},
  \citenamefont {Glawion}, \citenamefont {Walde}, \citenamefont {Pfaff},
  \citenamefont {Hollmark}, \citenamefont {Duda}, \citenamefont {Paetel},
  \citenamefont {Richter}, \citenamefont {Mannhart}, \citenamefont {Sing},\
  and\ \citenamefont {Claessen}}]{Berner2010}%
  \BibitemOpen
  \bibfield  {author} {\bibinfo {author} {\bibfnamefont {G.}~\bibnamefont
  {Berner}}, \bibinfo {author} {\bibfnamefont {S.}~\bibnamefont {Glawion}},
  \bibinfo {author} {\bibfnamefont {J.}~\bibnamefont {Walde}}, \bibinfo
  {author} {\bibfnamefont {F.}~\bibnamefont {Pfaff}}, \bibinfo {author}
  {\bibfnamefont {H.}~\bibnamefont {Hollmark}}, \bibinfo {author}
  {\bibfnamefont {L.-C.}\ \bibnamefont {Duda}}, \bibinfo {author}
  {\bibfnamefont {S.}~\bibnamefont {Paetel}}, \bibinfo {author} {\bibfnamefont
  {C.}~\bibnamefont {Richter}}, \bibinfo {author} {\bibfnamefont
  {J.}~\bibnamefont {Mannhart}}, \bibinfo {author} {\bibfnamefont
  {M.}~\bibnamefont {Sing}}, \ and\ \bibinfo {author} {\bibfnamefont
  {R.}~\bibnamefont {Claessen}},\ }\href@noop {} {\bibfield  {journal}
  {\bibinfo  {journal} {Phys. Rev. B.}\ }\textbf {\bibinfo {volume} {82}},\
  \bibinfo {pages} {241405(R)} (\bibinfo {year} {2010})}\BibitemShut {NoStop}%
\bibitem [{\citenamefont {Zhou}\ \emph {et~al.}(2011)\citenamefont {Zhou},
  \citenamefont {Radovic}, \citenamefont {Schlappa}, \citenamefont {Strocov},
  \citenamefont {Frison}, \citenamefont {Mesot}, \citenamefont {Patthey},\ and\
  \citenamefont {Schmitt}}]{Zhou2011}%
  \BibitemOpen
  \bibfield  {author} {\bibinfo {author} {\bibfnamefont {K.-J.}\ \bibnamefont
  {Zhou}}, \bibinfo {author} {\bibfnamefont {M.}~\bibnamefont {Radovic}},
  \bibinfo {author} {\bibfnamefont {J.}~\bibnamefont {Schlappa}}, \bibinfo
  {author} {\bibfnamefont {V.}~\bibnamefont {Strocov}}, \bibinfo {author}
  {\bibfnamefont {R.}~\bibnamefont {Frison}}, \bibinfo {author} {\bibfnamefont
  {J.}~\bibnamefont {Mesot}}, \bibinfo {author} {\bibfnamefont
  {L.}~\bibnamefont {Patthey}}, \ and\ \bibinfo {author} {\bibfnamefont
  {T.}~\bibnamefont {Schmitt}},\ }\href@noop {} {\bibfield  {journal} {\bibinfo
   {journal} {Phys. Rev. B.}\ }\textbf {\bibinfo {volume} {83}},\ \bibinfo
  {pages} {201402 (R)} (\bibinfo {year} {2011})}\BibitemShut {NoStop}%
\bibitem [{\citenamefont {Harada}\ \emph {et~al.}(2012)\citenamefont {Harada},
  \citenamefont {Kobayashi}, \citenamefont {Niwa}, \citenamefont {Senba},
  \citenamefont {Ohashi}, \citenamefont {Tokushima}, \citenamefont {Horikawa},
  \citenamefont {Shin},\ and\ \citenamefont {Oshima}}]{Harada2012}%
  \BibitemOpen
  \bibfield  {author} {\bibinfo {author} {\bibfnamefont {Y.}~\bibnamefont
  {Harada}}, \bibinfo {author} {\bibfnamefont {M.}~\bibnamefont {Kobayashi}},
  \bibinfo {author} {\bibfnamefont {H.}~\bibnamefont {Niwa}}, \bibinfo {author}
  {\bibfnamefont {Y.}~\bibnamefont {Senba}}, \bibinfo {author} {\bibfnamefont
  {H.}~\bibnamefont {Ohashi}}, \bibinfo {author} {\bibfnamefont
  {T.}~\bibnamefont {Tokushima}}, \bibinfo {author} {\bibfnamefont
  {Y.}~\bibnamefont {Horikawa}}, \bibinfo {author} {\bibfnamefont
  {S.}~\bibnamefont {Shin}}, \ and\ \bibinfo {author} {\bibfnamefont
  {M.}~\bibnamefont {Oshima}},\ }\href@noop {} {\bibfield  {journal} {\bibinfo
  {journal} {Rev. Sci. Instrum.}\ }\textbf {\bibinfo {volume} {83}},\ \bibinfo
  {pages} {013116} (\bibinfo {year} {2012})}\BibitemShut {NoStop}%
\bibitem [{\citenamefont {Walker}\ \emph {et~al.}(2014)\citenamefont {Walker},
  \citenamefont {de~la Torre}, \citenamefont {Bruno}, \citenamefont {Tamain},
  \citenamefont {Kim}, \citenamefont {Hoesch}, \citenamefont {Shi},
  \citenamefont {Bahramy}, \citenamefont {King},\ and\ \citenamefont
  {Baumberger}}]{Walker2014}%
  \BibitemOpen
  \bibfield  {author} {\bibinfo {author} {\bibfnamefont {S.~M.}\ \bibnamefont
  {Walker}}, \bibinfo {author} {\bibfnamefont {A.}~\bibnamefont {de~la Torre}},
  \bibinfo {author} {\bibfnamefont {F.~Y.}\ \bibnamefont {Bruno}}, \bibinfo
  {author} {\bibfnamefont {A.}~\bibnamefont {Tamain}}, \bibinfo {author}
  {\bibfnamefont {T.~K.}\ \bibnamefont {Kim}}, \bibinfo {author} {\bibfnamefont
  {M.}~\bibnamefont {Hoesch}}, \bibinfo {author} {\bibfnamefont
  {M.}~\bibnamefont {Shi}}, \bibinfo {author} {\bibfnamefont {M.~S.}\
  \bibnamefont {Bahramy}}, \bibinfo {author} {\bibfnamefont {P.~D.~C.}\
  \bibnamefont {King}}, \ and\ \bibinfo {author} {\bibfnamefont
  {F.}~\bibnamefont {Baumberger}},\ }\href@noop {} {\bibfield  {journal}
  {\bibinfo  {journal} {Phys. Rev. Lett.}\ }\textbf {\bibinfo {volume} {113}},\
  \bibinfo {pages} {177601} (\bibinfo {year} {2014})}\BibitemShut {NoStop}%
\bibitem [{\citenamefont {Ulrich}\ \emph {et~al.}(2008)\citenamefont {Ulrich},
  \citenamefont {Ghiringhelli}, \citenamefont {Piazzalunga}, \citenamefont
  {Braicovich}, \citenamefont {Brookes}, \citenamefont {Roth}, \citenamefont
  {Lorenz},\ and\ \citenamefont {Keimer}}]{Ulrich2008}%
  \BibitemOpen
  \bibfield  {author} {\bibinfo {author} {\bibfnamefont {C.}~\bibnamefont
  {Ulrich}}, \bibinfo {author} {\bibfnamefont {G.}~\bibnamefont
  {Ghiringhelli}}, \bibinfo {author} {\bibfnamefont {A.}~\bibnamefont
  {Piazzalunga}}, \bibinfo {author} {\bibfnamefont {L.}~\bibnamefont
  {Braicovich}}, \bibinfo {author} {\bibfnamefont {N.~B.}\ \bibnamefont
  {Brookes}}, \bibinfo {author} {\bibfnamefont {H.}~\bibnamefont {Roth}},
  \bibinfo {author} {\bibfnamefont {T.}~\bibnamefont {Lorenz}}, \ and\ \bibinfo
  {author} {\bibfnamefont {B.}~\bibnamefont {Keimer}},\ }\href@noop {}
  {\bibfield  {journal} {\bibinfo  {journal} {Phys. Rev. B.}\ }\textbf
  {\bibinfo {volume} {77}},\ \bibinfo {pages} {113102} (\bibinfo {year}
  {2008})}\BibitemShut {NoStop}%
\bibitem [{\citenamefont {Higuchi}\ \emph {et~al.}(1999)\citenamefont
  {Higuchi}, \citenamefont {Tsukamoto}, \citenamefont {Watanabe}, \citenamefont
  {Grush}, \citenamefont {Callcott}, \citenamefont {Perera}, \citenamefont
  {Ederer}, \citenamefont {Tokura}, \citenamefont {Harada}, \citenamefont
  {Tezuka},\ and\ \citenamefont {Shin}}]{Higuchi1999}%
  \BibitemOpen
  \bibfield  {author} {\bibinfo {author} {\bibfnamefont {T.}~\bibnamefont
  {Higuchi}}, \bibinfo {author} {\bibfnamefont {T.}~\bibnamefont {Tsukamoto}},
  \bibinfo {author} {\bibfnamefont {M.}~\bibnamefont {Watanabe}}, \bibinfo
  {author} {\bibfnamefont {M.~M.}\ \bibnamefont {Grush}}, \bibinfo {author}
  {\bibfnamefont {T.~A.}\ \bibnamefont {Callcott}}, \bibinfo {author}
  {\bibfnamefont {R.~C.}\ \bibnamefont {Perera}}, \bibinfo {author}
  {\bibfnamefont {D.~L.}\ \bibnamefont {Ederer}}, \bibinfo {author}
  {\bibfnamefont {Y.}~\bibnamefont {Tokura}}, \bibinfo {author} {\bibfnamefont
  {Y.}~\bibnamefont {Harada}}, \bibinfo {author} {\bibfnamefont
  {Y.}~\bibnamefont {Tezuka}}, \ and\ \bibinfo {author} {\bibfnamefont
  {S.}~\bibnamefont {Shin}},\ }\href@noop {} {\bibfield  {journal} {\bibinfo
  {journal} {Phys. Rev. B.}\ }\textbf {\bibinfo {volume} {60}},\ \bibinfo
  {pages} {7711} (\bibinfo {year} {1999})}\BibitemShut {NoStop}%
\bibitem [{\citenamefont {Strempfer}\ \emph {et~al.}(2013)\citenamefont
  {Strempfer}, \citenamefont {Francoual}, \citenamefont {Reuther},
  \citenamefont {Shukla}, \citenamefont {Skaugen}, \citenamefont
  {Schulte-Schrepping}, \citenamefont {Kracht},\ and\ \citenamefont
  {Franz}}]{Strempfer2013}%
  \BibitemOpen
  \bibfield  {author} {\bibinfo {author} {\bibfnamefont {J.}~\bibnamefont
  {Strempfer}}, \bibinfo {author} {\bibfnamefont {S.}~\bibnamefont
  {Francoual}}, \bibinfo {author} {\bibfnamefont {D.}~\bibnamefont {Reuther}},
  \bibinfo {author} {\bibfnamefont {D.~K.}\ \bibnamefont {Shukla}}, \bibinfo
  {author} {\bibfnamefont {A.}~\bibnamefont {Skaugen}}, \bibinfo {author}
  {\bibfnamefont {H.}~\bibnamefont {Schulte-Schrepping}}, \bibinfo {author}
  {\bibfnamefont {T.}~\bibnamefont {Kracht}}, \ and\ \bibinfo {author}
  {\bibfnamefont {H.}~\bibnamefont {Franz}},\ }\href@noop {} {\bibfield
  {journal} {\bibinfo  {journal} {Journal of Synchrotron Radiation}\ }\textbf
  {\bibinfo {volume} {20}},\ \bibinfo {pages} {541} (\bibinfo {year}
  {2013})}\BibitemShut {NoStop}%
\bibitem [{\citenamefont {Yoshii}\ \emph {et~al.}(2012)\citenamefont {Yoshii},
  \citenamefont {Jarrige}, \citenamefont {Suzuki}, \citenamefont {Matsumura},
  \citenamefont {Nishihata}, \citenamefont {Yoneda}, \citenamefont {Fukuda},
  \citenamefont {Tamura}, \citenamefont {Ito}, \citenamefont {Mukoyama},
  \citenamefont {Tochio}, \citenamefont {Shinotruka},\ and\ \citenamefont
  {Fukushima}}]{Yoshii2012}%
  \BibitemOpen
  \bibfield  {author} {\bibinfo {author} {\bibfnamefont {K.}~\bibnamefont
  {Yoshii}}, \bibinfo {author} {\bibfnamefont {I.}~\bibnamefont {Jarrige}},
  \bibinfo {author} {\bibfnamefont {C.}~\bibnamefont {Suzuki}}, \bibinfo
  {author} {\bibfnamefont {D.}~\bibnamefont {Matsumura}}, \bibinfo {author}
  {\bibfnamefont {Y.}~\bibnamefont {Nishihata}}, \bibinfo {author}
  {\bibfnamefont {Y.}~\bibnamefont {Yoneda}}, \bibinfo {author} {\bibfnamefont
  {T.}~\bibnamefont {Fukuda}}, \bibinfo {author} {\bibfnamefont
  {K.}~\bibnamefont {Tamura}}, \bibinfo {author} {\bibfnamefont
  {Y.}~\bibnamefont {Ito}}, \bibinfo {author} {\bibfnamefont {T.}~\bibnamefont
  {Mukoyama}}, \bibinfo {author} {\bibfnamefont {T.}~\bibnamefont {Tochio}},
  \bibinfo {author} {\bibfnamefont {H.}~\bibnamefont {Shinotruka}}, \ and\
  \bibinfo {author} {\bibfnamefont {S.}~\bibnamefont {Fukushima}},\ }\href@noop
  {} {\bibfield  {journal} {\bibinfo  {journal} {J. Phys. Chem. Solids}\
  }\textbf {\bibinfo {volume} {73}},\ \bibinfo {pages} {1106} (\bibinfo {year}
  {2012})}\BibitemShut {NoStop}%
\bibitem [{\citenamefont {Takizawa}\ \emph {et~al.}(2011)\citenamefont
  {Takizawa}, \citenamefont {Tsuda}, \citenamefont {Susaki}, \citenamefont
  {Hwang},\ and\ \citenamefont {Fujimori}}]{Takizawa2011}%
  \BibitemOpen
  \bibfield  {author} {\bibinfo {author} {\bibfnamefont {M.}~\bibnamefont
  {Takizawa}}, \bibinfo {author} {\bibfnamefont {S.}~\bibnamefont {Tsuda}},
  \bibinfo {author} {\bibfnamefont {T.}~\bibnamefont {Susaki}}, \bibinfo
  {author} {\bibfnamefont {H.~Y.}\ \bibnamefont {Hwang}}, \ and\ \bibinfo
  {author} {\bibfnamefont {A.}~\bibnamefont {Fujimori}},\ }\href@noop {}
  {\bibfield  {journal} {\bibinfo  {journal} {Phys. Rev. B.}\ }\textbf
  {\bibinfo {volume} {84}},\ \bibinfo {pages} {245124} (\bibinfo {year}
  {2011})}\BibitemShut {NoStop}%
\bibitem [{\citenamefont {Salluzzo}\ \emph {et~al.}(2013)\citenamefont
  {Salluzzo}, \citenamefont {Gariglio}, \citenamefont {Torrelles},
  \citenamefont {Ristic}, \citenamefont {Capua}, \citenamefont {Drnec},
  \citenamefont {Sala}, \citenamefont {Ghiringhelli}, \citenamefont {Felici},\
  and\ \citenamefont {Brookes}}]{Salluzzo2013}%
  \BibitemOpen
  \bibfield  {author} {\bibinfo {author} {\bibfnamefont {M.}~\bibnamefont
  {Salluzzo}}, \bibinfo {author} {\bibfnamefont {S.}~\bibnamefont {Gariglio}},
  \bibinfo {author} {\bibfnamefont {X.}~\bibnamefont {Torrelles}}, \bibinfo
  {author} {\bibfnamefont {Z.}~\bibnamefont {Ristic}}, \bibinfo {author}
  {\bibfnamefont {R.~D.}\ \bibnamefont {Capua}}, \bibinfo {author}
  {\bibfnamefont {J.}~\bibnamefont {Drnec}}, \bibinfo {author} {\bibfnamefont
  {M.~M.}\ \bibnamefont {Sala}}, \bibinfo {author} {\bibfnamefont
  {G.}~\bibnamefont {Ghiringhelli}}, \bibinfo {author} {\bibfnamefont
  {R.}~\bibnamefont {Felici}}, \ and\ \bibinfo {author} {\bibfnamefont {N.~B.}\
  \bibnamefont {Brookes}},\ }\href@noop {} {\bibfield  {journal} {\bibinfo
  {journal} {Adv. Mater.}\ }\textbf {\bibinfo {volume} {25}},\ \bibinfo {pages}
  {2333} (\bibinfo {year} {2013})}\BibitemShut {NoStop}%
\bibitem [{\citenamefont {j.~Son}\ \emph {et~al.}(2009)\citenamefont {j.~Son},
  \citenamefont {Cho}, \citenamefont {Lee}, \citenamefont {Lee},\ and\
  \citenamefont {Han}}]{Son2009}%
  \BibitemOpen
  \bibfield  {author} {\bibinfo {author} {\bibfnamefont {W.}~\bibnamefont
  {j.~Son}}, \bibinfo {author} {\bibfnamefont {E.}~\bibnamefont {Cho}},
  \bibinfo {author} {\bibfnamefont {B.}~\bibnamefont {Lee}}, \bibinfo {author}
  {\bibfnamefont {J.}~\bibnamefont {Lee}}, \ and\ \bibinfo {author}
  {\bibfnamefont {S.}~\bibnamefont {Han}},\ }\href@noop {} {\bibfield
  {journal} {\bibinfo  {journal} {Phys. Rev. B.}\ }\textbf {\bibinfo {volume}
  {79}},\ \bibinfo {pages} {245411} (\bibinfo {year} {2009})}\BibitemShut
  {NoStop}%
\bibitem [{\citenamefont {Salluzzo}\ \emph {et~al.}(2009)\citenamefont
  {Salluzzo}, \citenamefont {Cezar}, \citenamefont {Brookes}, \citenamefont
  {Bisogni}, \citenamefont {Luca}, \citenamefont {Richter}, \citenamefont
  {Thiel}, \citenamefont {Mannhart}, \citenamefont {Huijben}, \citenamefont
  {Brinkman}, \citenamefont {Rijnders},\ and\ \citenamefont
  {Ghiringhelli}}]{Salluzzo2009}%
  \BibitemOpen
  \bibfield  {author} {\bibinfo {author} {\bibfnamefont {M.}~\bibnamefont
  {Salluzzo}}, \bibinfo {author} {\bibfnamefont {J.~C.}\ \bibnamefont {Cezar}},
  \bibinfo {author} {\bibfnamefont {N.~B.}\ \bibnamefont {Brookes}}, \bibinfo
  {author} {\bibfnamefont {V.}~\bibnamefont {Bisogni}}, \bibinfo {author}
  {\bibfnamefont {G.~M.~D.}\ \bibnamefont {Luca}}, \bibinfo {author}
  {\bibfnamefont {C.}~\bibnamefont {Richter}}, \bibinfo {author} {\bibfnamefont
  {S.}~\bibnamefont {Thiel}}, \bibinfo {author} {\bibfnamefont
  {J.}~\bibnamefont {Mannhart}}, \bibinfo {author} {\bibfnamefont
  {M.}~\bibnamefont {Huijben}}, \bibinfo {author} {\bibfnamefont
  {A.}~\bibnamefont {Brinkman}}, \bibinfo {author} {\bibfnamefont
  {G.}~\bibnamefont {Rijnders}}, \ and\ \bibinfo {author} {\bibfnamefont
  {G.}~\bibnamefont {Ghiringhelli}},\ }\href@noop {} {\bibfield  {journal}
  {\bibinfo  {journal} {Phys. Rev. Lett.}\ }\textbf {\bibinfo {volume} {102}},\
  \bibinfo {pages} {166804} (\bibinfo {year} {2009})}\BibitemShut {NoStop}%
\bibitem [{\citenamefont {Lin}\ and\ \citenamefont {Demkov}(2013)}]{Lin2013}%
  \BibitemOpen
  \bibfield  {author} {\bibinfo {author} {\bibfnamefont {C.}~\bibnamefont
  {Lin}}\ and\ \bibinfo {author} {\bibfnamefont {A.~A.}\ \bibnamefont
  {Demkov}},\ }\href@noop {} {\bibfield  {journal} {\bibinfo  {journal} {Phys.
  Rev. Lett.}\ }\textbf {\bibinfo {volume} {111}},\ \bibinfo {pages} {217601}
  (\bibinfo {year} {2013})}\BibitemShut {NoStop}%
\bibitem [{\citenamefont {Jeschke}\ \emph {et~al.}(2015)\citenamefont
  {Jeschke}, \citenamefont {Shen},\ and\ \citenamefont
  {Valenti}}]{Jeschke2015}%
  \BibitemOpen
  \bibfield  {author} {\bibinfo {author} {\bibfnamefont {H.~O.}\ \bibnamefont
  {Jeschke}}, \bibinfo {author} {\bibfnamefont {J.}~\bibnamefont {Shen}}, \
  and\ \bibinfo {author} {\bibfnamefont {R.}~\bibnamefont {Valenti}},\
  }\href@noop {} {\bibfield  {journal} {\bibinfo  {journal} {New J. Phys.}\
  }\textbf {\bibinfo {volume} {17}},\ \bibinfo {pages} {023034} (\bibinfo
  {year} {2015})}\BibitemShut {NoStop}%
\bibitem [{\citenamefont {Ide}\ and\ \citenamefont {Kotani}(1998)}]{Ide1998}%
  \BibitemOpen
  \bibfield  {author} {\bibinfo {author} {\bibfnamefont {T.}~\bibnamefont
  {Ide}}\ and\ \bibinfo {author} {\bibfnamefont {A.}~\bibnamefont {Kotani}},\
  }\href@noop {} {\bibfield  {journal} {\bibinfo  {journal} {J. Phys. Soc.
  Jpn.}\ }\textbf {\bibinfo {volume} {67}},\ \bibinfo {pages} {3621} (\bibinfo
  {year} {1998})}\BibitemShut {NoStop}%
\bibitem [{\citenamefont {Ide}\ and\ \citenamefont {Kotani}(2000)}]{Ide2000}%
  \BibitemOpen
  \bibfield  {author} {\bibinfo {author} {\bibfnamefont {T.}~\bibnamefont
  {Ide}}\ and\ \bibinfo {author} {\bibfnamefont {A.}~\bibnamefont {Kotani}},\
  }\href@noop {} {\bibfield  {journal} {\bibinfo  {journal} {J. Phys. Soc.
  Jpn.}\ }\textbf {\bibinfo {volume} {69}},\ \bibinfo {pages} {1895} (\bibinfo
  {year} {2000})}\BibitemShut {NoStop}%
\end{thebibliography}
\end{document}